\documentclass[journal,10pt]{IEEEtran}
\usepackage{multirow}

\usepackage{stackengine}
\usepackage{amsmath,graphicx}

\usepackage[sort]{cite}
\usepackage{subfig}

 \usepackage{algorithm}
\usepackage{algpseudocode}
\usepackage{color,soul}

\ifCLASSINFOpdf
\else
\fi
\begin{document}
%
\title{Adaptive dictionary based approach for background noise and speaker classification and subsequent  source separation }

%
%
%

\author{K V Vijay Girish$^+$, A G Ramakrishnan$^+$,~\IEEEmembership{Senior Member,~IEEE} and T V Ananthapadmanabha$^*$. \\
        $^+$Department of Electrical Engineering,
               Indian Institute of Science, Bangalore\\
                $^*$Voice and Speech Systems, Malleswaram, Bangalore, India \\}

%
%

\markboth{Journal of \LaTeX\ Class Files,~Vol.~14, No.~8, August~2015}%
{Shell \MakeLowercase{\textit{et al.}}: Bare Demo of IEEEtran.cls for IEEE Journals}
%



\maketitle

\begin{abstract}

  A judicious combination of dictionary learning methods, block sparsity and source recovery algorithm are used in a hierarchical manner to identify the noises and the speakers from a noisy conversation between two people. Conversations are simulated using speech from two speakers, each with a different background noise, with varied SNR values, down to -10 dB. Ten each of randomly chosen male and female speakers from the TIMIT database and all the noise sources from the NOISEX database are used for the simulations. For speaker identification, the relative value of weights recovered   is used to select an appropriately small subset of the test data, assumed to contain speech. This novel choice of using varied amounts of test data results in an improvement in the speaker recognition rate of around 15\% at  SNR of 0 dB. Speech and noise are separated using dictionaries of the estimated speaker and noise, and an improvement of signal to distortion ratios of up to 10\% is achieved at SNR of 0 dB. K-medoid and cosine similarity based dictionary learning methods lead to better recognition of the background noise and the speaker. Experiments are also conducted on cases, where either the background noise or the speaker is outside the set of trained dictionaries. In such cases, adaptive dictionary learning leads to performance comparable to the other case of complete dictionaries.
  
\end{abstract}


%
\IEEEpeerreviewmaketitle

\section{Introduction}
%
%
%
%

\subsection{Motivation}

Audio signals occurring in nature which  are generally of interest to us are a mixture of foreground speech and background noise like factory or babble noise. Analysis of these audio signals is useful for acquiring information about the speaker, background noise environment, location of speech segments and source separation.
Estimation of background noise helps us to narrow down to the possible geographical location of the speaker. Other applications include selection of appropriate noise model for speech enhancement. Speaker estimation is useful for tracking the person and selecting an appropriate speaker model for source separation. 

 Estimation of frame-wise energy of speech source is used for  speech segment detection. These speech segments can then be processed for further speech analysis and recognition.  We also estimate the segmental SNR and detect the speech segments in a noisy speech signal.

In the case of when both speaker and background noise are unknown, the noise can be mapped to the nearest noise and speaker index, and the model for the same can be used for separation or enhancement. The model corresponding  to the nearest noise/ speaker index can be adapted using  the segments  of the test signal containing noise/ speech only segments. SNR estimation and location of speech segments gives us the temporal information of speech/ noise only segments  occurring within an audio signal. So, analysis of audio signals can be done even though the noise/ speaker components belong to a unknown set.   Identification of background environment and speaker is useful in other applications like  hearing aids  \cite{turner}, forensics \cite{ikram} and  robotic navigation systems \cite{chu1}.     In this work, we  address the problem of classification and separation of a mixed audio signal containing multiple speakers and noises using  the concept of dictionary based representation, block sparsity \cite{eldar} and sparse non-negative recovery \cite{virtanen}. The advantage of using dictionary based approach for classification is that sparse representation using dictionaries can assume that the signal to be classified may be mixed with noises whose dictionary is known or can be estimated. However, conventional methods for classification generally fail when the signal to be classified is mixed with noises at low signal to noise ratios.

 We propose a novel, rule-based, sparse representation  approach to first identify the type of noises and speakers present in a  mixed audio signal and subsequently separate the speech and noise signals. We assume the mixed audio signal consists of speech from two speakers, each in the presence of a specific type of noise.  The objective score of signal to distortion ratio and improvement in signal to noise ratio (SNR) shows the speech enhancement achieved.

  \subsection{Literature review}
Work on audio content analysis and scene classification has been done by many researchers. Lu et. al. \cite{lu}   used K-nearest neighbor and line spectral pairs-vector quantization to classify audio into speech, music, environmental sound and silence. Zhang et.al. \cite{zhang} classified and segmented audio signal from movies or  TV programs using simple audio features like  energy function and average zero crossing rate. 
  A review of the state of the art in acoustic scene classification was done by Barchiesi et.al. \cite{barchesi} while  Giannoulis et al. \cite{giannoulis} evaluated 11 algorithms  along with a baseline system for scene classification using a statistical model or majority vote based classifier. Cauchi \cite{cauchi} did  auditory scene classification using non-negative matrix factorization (NMF).
  
Lyon \cite{lyon} explored   machinery noise diagnostics  while Shirkhodaie et. al. \cite{shirkhodaie} surveyed  acoustic signature classification  of aircrafts or vehicles.  Kates \cite{kates} classified noise for hearing aid applications based on variation of signal envelope as features.   Classification of different kinds of noise and speech using line spectral frequencies as features was done by  Maleh et. al. \cite{maleh}. A system using a hidden Markov model classifier and  log-spectral features  to classify twenty different types of sounds was devised by  Casey \cite{casey}.  Matching pursuit based features were combined with mel-frequency cepstral coefficients by Chu et. al. \cite{chu} to recognize 14 different environmental sounds. Techniques for  stationary and non-stationary environmental sound recognition was surveyed by Chachada et. al. \cite{chachada}.   Malik \cite{malik} estimated the amount of reverberation and background noise variance using a statistical technique.

  Sparsity based speaker identification using discriminative dictionary learning was done by Tzagkarakis et. al. \cite{tzagkarakis} while  non-negative matrix factorization for feature extraction was explored by  Joder et. al. \cite{joder}. 
    
   Machine listening research to solve real world problems in perceptual computing was explored by  Malkin \cite{malkin}. Our paper addresses some components of machine listening like classification and separation of multiple speakers and noises in a mixed audio signal.
  

  A   dictionary is a collection of vectors called as atoms learnt  from the feature vectors of a  large training data. Given a test feature vector expressed as a linear combination of these atoms, source recovery is the method of  estimating weights corresponding to these atoms.
      
              Representation of audio signals   as a linear combination of non-negative dictionary atoms   is shown for audio source separation \cite{virtanen0,ozerov,mysore}, recognition \cite{bertin,gemmeke,raj}, classification \cite{cho,zubair} and coding \cite{nikunen,plumbley}.        
      Dictionary learning (DL) method by random selection of features from the training data is done in \cite{virtanen}. K-means clustering  has been used for DL by \cite{coates}. The relation between DL and vector quantization was shown by \cite{delgado} . A probabilistic model of the features has been used by Olshausen \cite{olshausen} and Lewicki \cite{lewicki} for DL . Engan et al. \cite{engan} performed DL using dictionary update method (minimization of mean square error) and sparse coding using  orthogonal matching pursuit (OMP) \cite{pati} or  focal underdetermined system solver (FOCUSS) \cite{gorodnitsky}. Recursive least squares dictionary learning (RLS-DLA) \cite{skretting}, K-SVD \cite{aharon}, simultaneous codeword optimisation (SimCO) \cite{dai} and fast dictionary learning \cite{jafari} are other DL algorithms.  DL and source recovery methods have been used for classification of objects in images by learning class-specific dictionaries \cite{kong}. Shafiee et al. \cite{shafiee} have used three different DL methods to classify faces and digits in images.

       Some source recovery algorithms are Matching pursuit \cite{mallat}, orthogonal matching pursuit (OMP) \cite{pati}, basis pursuit \cite{chen}, focal underdetermined system solver (FOCUSS) \cite{gorodnitsky}  and active-set Newton algorithm (ASNA) \cite{virtanen} .  DL and source recovery methods have been used for classification of objects in images by learning class-specific dictionaries \cite{kong}.

       
            We have used the active-set Newton algorithm (ASNA) \cite{virtanen} algorithm for source recovery in the testing phase. The benefits of this algorithm are that the it returns non-negative weights and can handle non-stationary signals like speech.  The training phase for the  classification problem is DL from various speaker/ noise sources where  different dictionary atoms encompass the variation in the spectral characteristics.
       
       Girish et. al. \cite{kv} classified the speaker and noise type using sparse representation in the case of a single speaker and noise. This paper deals with segments containing multiple noises and speakers,  and estimates all the speakers and noises embedded in the mixed audio signal. Also, this paper uses additional dictionary learning methods and separates speech and noise sources.  In addition, we deal with unknown speaker and noise sources using adaptive dictionaries.

  \subsection{Contributions}
    
        The major contributions of this paper are:  (1) Simulating audio signals containing a  concatenation of multiple noises, with speech from different speakers mixed with each type of noise, (2) Block sparsity and concatenated dictionary based classification of multiple speech and noise sources in a mixed audio signal, (3) A rule based  divide and conquer approach to segment and classify multiple noise segments, (4) Using high energy frames along with relatively higher average weights corresponding to speaker dictionaries with respect to that of noise dictionary for speaker classification (3) Exploring different dictionary learning methods for classification and separation of audio signals, (5) Adaptive update of noise and speaker dictionaries and a novel generalized algorithm to update dictionaries using noise and speech only parts of the noisy speech signal and evaluating improvement in performance.
%
%
%

\section{Dictionary learning methods}\label{dictlearn}
  A dictionary is defined as a matrix $\textbf{D}\in \rm I\!R^{P\times K}$ (with $P$ as the dimension of the acoustic feature vector) containing $K$ column vectors called atoms, denoted as $\mathbf{d}_k, 1\leq k\leq K$. Any real valued feature vector, $y$ can be represented as $\mathbf{y}\approx \textbf{D}\mathbf{x}$, where $\mathbf{x}\in \rm I\!R^K$ is the vector containing weights for each dictionary atom. The vector $\mathbf{x}$ is estimated by minimizing the distance $dist(\mathbf{y},\textbf{D}\mathbf{x})$, where $dist()$ is a distance metric between $\mathbf{y}$ and $\textbf{D}\mathbf{x}$ such as $L_2$ norm or Kullback-Leibler (KL)-divergence \cite{virtanen}. In case the dictionary $\textbf{D}$ is overcomplete, the weight vector $\textbf{x}$ tends to be sparse. 
  
  Dictionary learning is the method of constructing the dictionary $\textbf{D}$, given the training features for each source. All the atoms of the dictionary are normalized to unit $L_2$ norm. In this paper, we use dictionary learning methods which learn non-negative dictionary atoms, since non-negative training features are used and we want to avoid negative representation of features.
  
   Threshold dependent cosine similarity based dictionary learning (TDCS) proposed in \cite{kv1} is used apart from random selection and clustering based methods for dictionary learning.

   In TDCS, each dictionary atom  is selected such that it is as uncorrelated as possible to the rest of the atoms belonging to the same as well as other sources. The correlation between a pair of atoms $\mathbf{d}_n,\mathbf{d}_j$ is measured using the  cosine similarity   as:
   
   \begin{equation}
   cs(\mathbf{d}_n,\mathbf{d}_j)=\mathbf{d}_n^T\mathbf{d}_j/(||\mathbf{d}_n||||\mathbf{d}_j||)
   \end{equation}
   Two types of cosine similarity measures are used: (a) within-class cosine similarity (within-CS)  defined as  $cs_w(\mathbf{d}_n,\mathbf{d}_j),\;$ $ \mathbf{d}_n,\mathbf{d}_j\in \textbf{D}^k , n\neq j$
   where $\textbf{D}^k$ is the dictionary for a specific source; and (b) between-class cosine similarity (between-CS) defined as $cs_b(\mathbf{d}_n,\mathbf{d}_j),\; \mathbf{d}_n\in \textbf{D}^k,\,\mathbf{d}_j\in \textbf{D}^m, k\neq m $.
   
   For each source, the dictionary atoms  are learnt such that the cosine similarity between the atoms is below a set threshold, chosen based on the desired performance. A randomly selected  feature vector, denoted as $\mathbf{y}_r$, is taken as the first atom for the first source, $\mathbf{d}_1^1$ . The rest of the atoms are learnt by  random selection of the feature vectors (excluding features already selected as atoms): $t^{th}$ feature, $\mathbf{y}_t$, is selected as the $n^{th}$ atom, $\mathbf{d}^1_n$ of dictionary $\textbf{D}^1$ if the maximum of within-CS, $\underset{j}{\max} \;cs_{w}(\mathbf{y}_t,\mathbf{d}_j^1),\; j< n$ (similar to coherence in \cite{tropp}) is less than a threshold $T_w$.
   
   The selection of dictionary atoms is stopped once the number of dictionary atoms reaches a pre-decided number $N_{A}$. In case $N_{A}$ atoms are not obtained, additional features, which do not satisfy the within-class threshold $T_w$, are appended in the order of increasing $max \;cs_w$.
   
   To learn dictionaries for subsequent sources, atoms are learnt using an additional constraint: $\mathbf{y}_t$ is selected as the $n^{th}$ atom $\mathbf{d}_n^k$ for the $k^{th} $ dictionary $\textbf{D}^k$, if  $max\; cs_b(\mathbf{y}_t, \mathbf{d}_j^h),$ $ \mathbf{d}_j^h\in \textbf{D}^h,\:h<k, 1\leq j \leq N_A $  is less than a threshold $T_{b}$.
   
   The threshold  $T_w$  ensures that atoms within the same source dictionary are as uncorrelated as possible, while $T_b$ ensures that atoms from different source dictionaries are maximally uncorrelated. Lower the values of the thresholds $T_w$ and $T_b$, greater is the uncorrelatedness between the dictionary atoms.

     The TDCS algorithm is summarized in Algorithm \ref{dictalgoi}. For the sake of simplicity, the algorithm does not show the appending of additional dictionary atoms when $N_{A}$ atoms could not be obtained.
   
  \begin{algorithm}{Dictionary learning using TDCS}\label{dictalgoi}
  \begin{algorithmic}[1]
  \State  \textbf{Initialize:}	Dictionary index $k=1$;  $\textbf{D}^k=\mathbf{d}_1^1=\mathbf{y}_r$;  Atom index $n=2$; set $T_w$ and $T_b$.
  	
  	\Repeat
  	\State Extract $N$ number of  features denoted as $\mathbf{y}_t, 1\leq t\leq N$ from the $k^{th}$  source.
  	\Repeat
  	
  	\State If $n>1$, find the maximum of within-CS, $m_i$ as:
  	
  	 $\max( cs_w(\mathbf{y}_t,\mathbf{d}^k_j)\;\forall\: j=1...n-1)$
  	 	 
  \State	If $k>1$, find the maximum of between-CS, $ m_I$   as:
  		
  		 $\max(cs_b(\mathbf{y}_t,\mathbf{d}^h_j)\:\forall\: j=1..N_A,h< k)$
   \If{$m_i\leq T_w$  and  $m_I\leq T_b$ (for $k>1$)}
  \State Assign randomly selected $\mathbf{f}_t$ as the $n^{th}$ atom: $\mathbf{d}^k_n=\mathbf{y}_t$ and append to the dictionary: $\textbf{D}^k=[\textbf{D}^k\; \mathbf{d}^k_n]$
  \State $n=n+1$
  
  \EndIf
  
  \Until{$n>N_{A}$ }
  		
  	\State $k=k+1$; $n=1$
  \Until{All source dictionaries are learnt}
  \end{algorithmic}
  \end{algorithm}

  The following five dictionary learning methods have been used in this paper :

\begin{enumerate}
\item \textit{Random selection of features}: Features randomly picked up from the training set using  a uniform distribution  are assigned as the dictionary atoms.
\item \textit{K-means clustering}: K-means  clustering  \cite{spath} using Euclidian  distance  measure, where $K $ is the number of atoms and  normalized means are used as the dictionary  atoms.
\item \textit{K-medoid clustering }: This is performed using the algorithm proposed by Park et.al. \cite{park}  and the $K$ medoids are used as the dictionary atoms.
\item  \textit{TDCS-0.9}:  TDCS algorithm with the intra and between-class thresholds as $T_w=0.9,\: T_b=0.9$

\item  \textit{TDCS-0.8}:  TDCS algorithm with the intra and between-class thresholds as $T_w=0.8,\: T_b=0.8$

\end{enumerate}

\begin{figure}[!h]
       \centering
      \subfloat[Noises]{
       \includegraphics[width=.5\textwidth,height=.13\textheight]{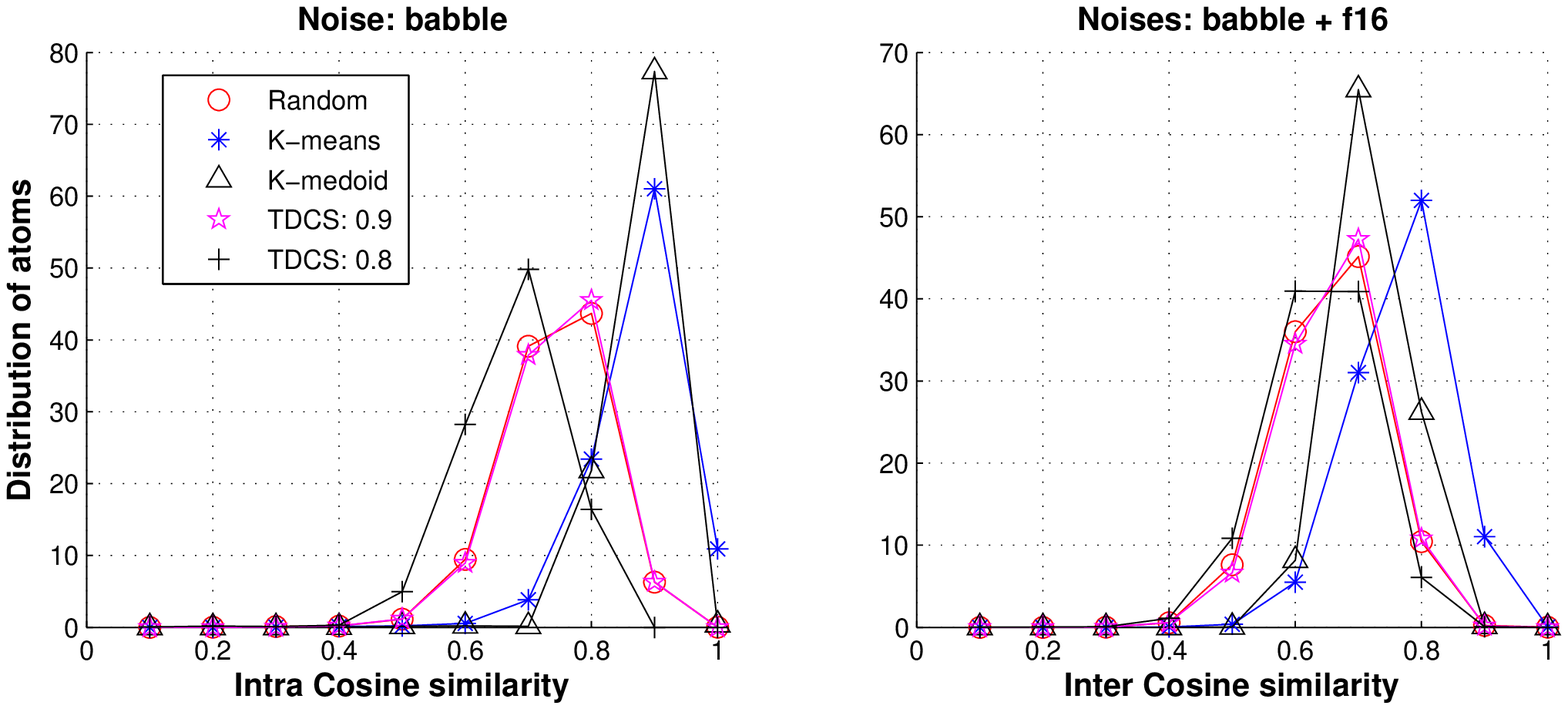}}
       
       \subfloat[Male speakers]{
              \includegraphics[width=.5\textwidth,height=.13\textheight]{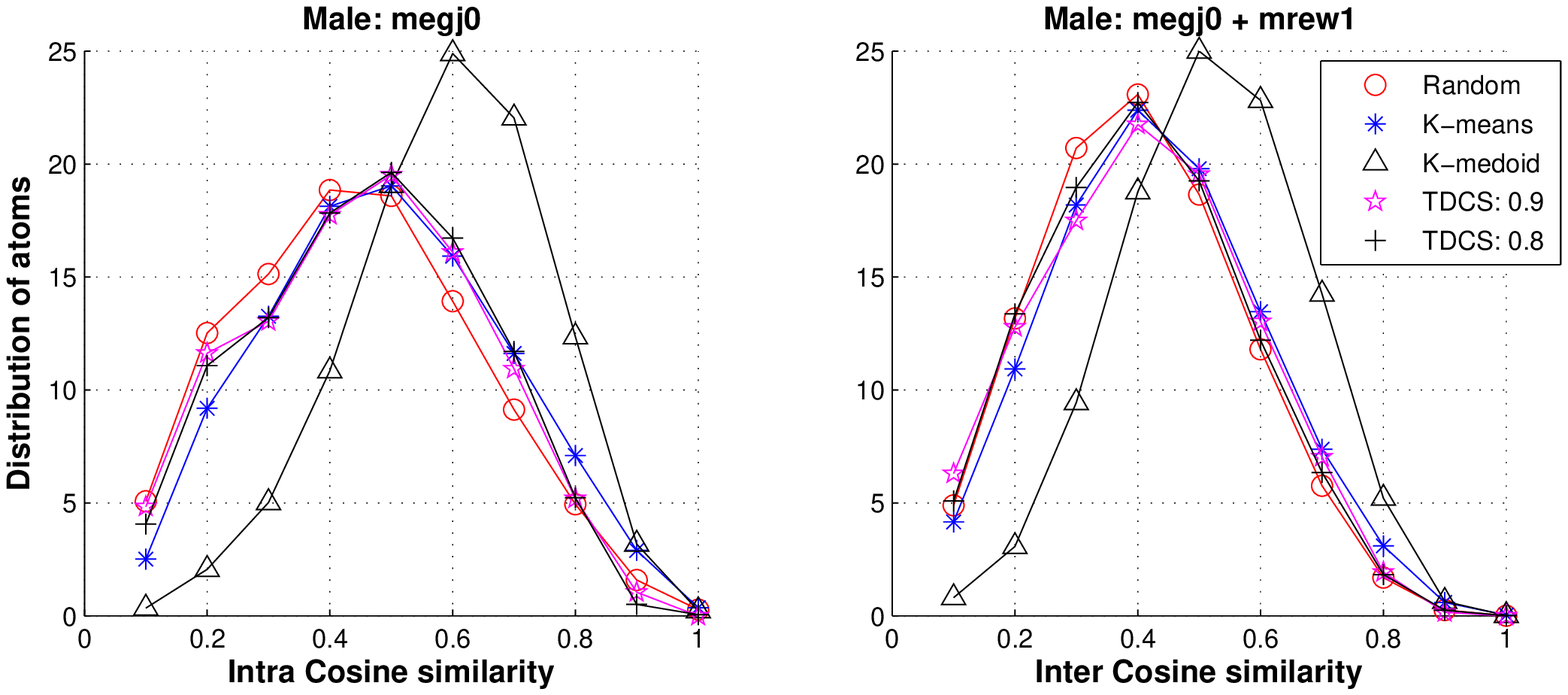}
       }
       
           \subfloat[Female speakers]{
                     \includegraphics[width=.5\textwidth,height=.13\textheight]{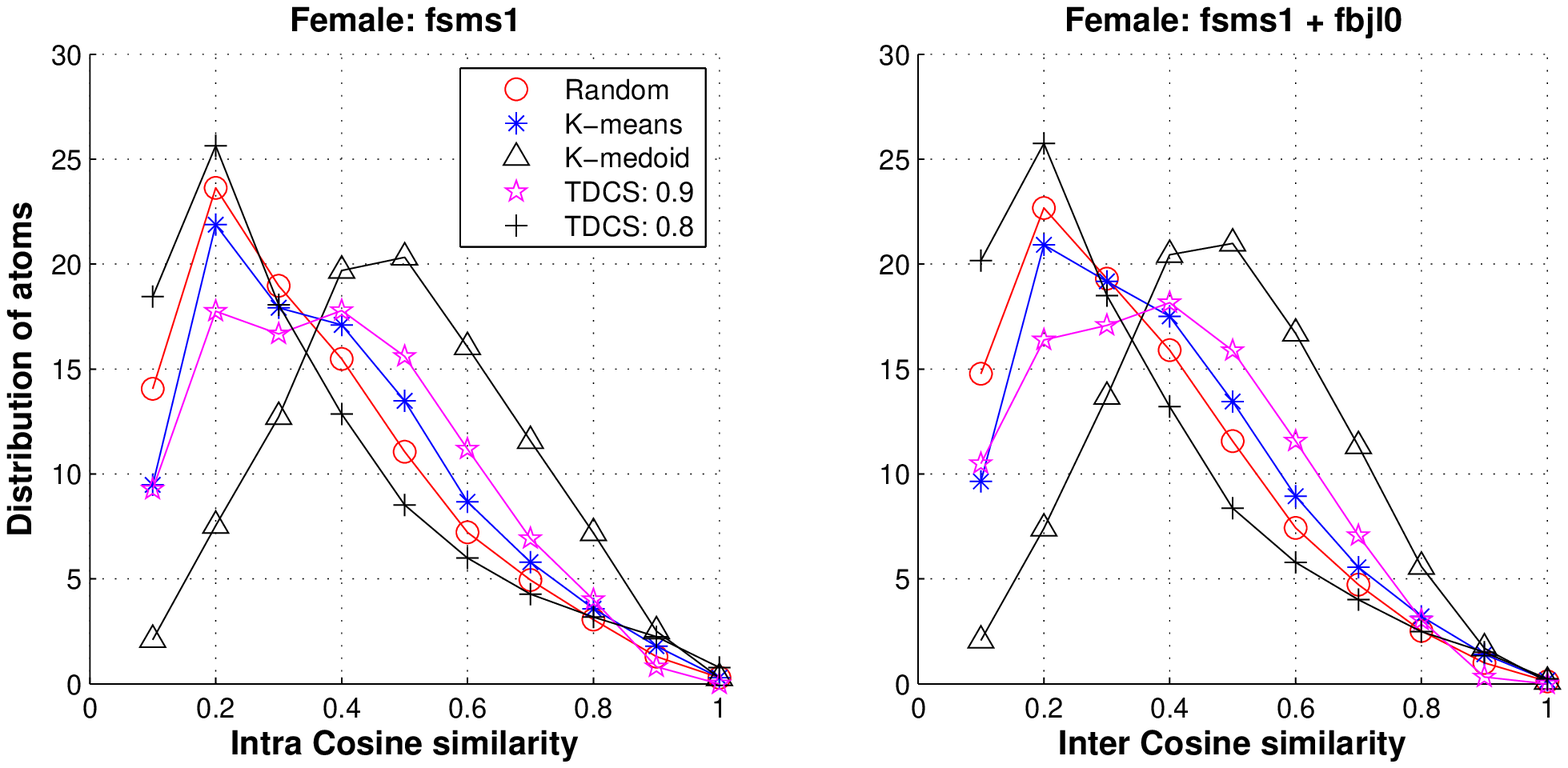}
              }
       \caption{Percentage distribution of number of dictionary atom combinations across within-class cosine similarity and between-class cosine similarity for (a) two noises; (b) two male speakers; and (c) two female speakers. }
       \label{csim}
       \end{figure}

Figure \ref{csim} shows the percentage distribution of number of dictionary atom combinations as a function of within-cosine similarity and between-cosine similarity. It is seen  in Fig. \ref{csim} (a,c) for noises and female speakers that TDCS using a threshold of $T_w=0.8,\;,T_b=0.8$ has higher number of atoms combinations with low cosine similarity.

%
%
%
%

\section{Problem formulation}

The test time domain noisy signal, $s[n]$ is simulated as a concatenation of two noisy speech signals, $s_1[n]$ and $s_2[n]$. Each $s_i[n], i\in 1,2$ is simulated as a linear combination of a speech,  $s_{sp}[n]$ and a noise source, $s_{ns}[n]$  as

\begin{equation}
s_i[n]=s_{sp}[n]+s_{ns}[n]
\end{equation}
Both the speaker and noise sources for the two noisy speech signals are different and are constrained to belong to a specific set of speaker and noise sources. So, $s[n]$ contains two different speakers and noise sources, each noise segment containing speech utterance by a single speaker. The first noise segment contains a male speaker and second noise segment contains a female speaker. The common occurrence of these type of test audio signals are telephonic conversations where the two speakers are in a different noise environment.

\begin{figure}[!h]
      \centering
     
      \includegraphics[width=.5\textwidth,height=.32\textheight]{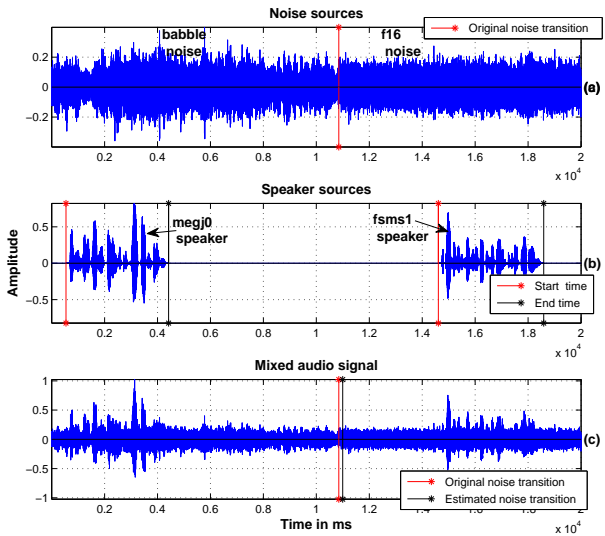}
      
      \caption{Illustration of two noise sources, two speaker sources and the mixed audio  signal at a SNR of 0 dB. }
      \label{spnsch}
      \end{figure}

Figure \ref{spnsch} shows the concatenated noise sources in (a), concatenated clean speech utterances from two  different speakers in (b) and the mixed audio signal as a linear combination of (a) and (b) at a SNR of 0 dB. The transition from  babble noise to f16 noise and the boundaries of the speech segments are  depicted in the figure.

Given the mixed audio signal, we find the  instant of transition from one noise to another and classify the noises. Within each noise segment, we estimate the speaker source and separate the speech from noise. We  also estimate the possible regions containing the speech segments. The approach and the algorithms used for the same are described in the following subsections. 

\subsection{Feature extraction}
 Frames of 60 ms duration are extracted with a shift of 15 ms from the training set of speaker and noise sources.  Features are extracted as the magnitude of the short-time Fourier transform  of these frames using a Hanning window. For dictionary learning, features having very low relative energy (0.001 times) compared to the average energy of the features are removed.  It is to be noted that all the features and dictionary atoms are non-negative as we require a non-negative representation during classification and separation stage.  Dictionaries for all the sources are learnt separately by the five dictionary learning methods as given in Sec. \ref{dictlearn} using  $K (=500)$ number of atoms. Features for each speaker and noise source are extracted separately and the corresponding dictionaries are built. The dictionaries for $N_{sp}$ speaker and  $N_{ns}$  noise sources are denoted as $\textbf{D}_{sp}^i, \;1\leq i\leq N_{sp}$  and  $\textbf{D}_{ns}^i, \;1\leq i\leq N_{ns}$, respectively.

\subsection{Noise segmentation and classification stage}

In a test audio signal, the class of noise source is constant across various frames over a significant span of time as the  noise class may not change unless the speaker is traveling. So, we follow a top down approach of classifying and segmenting a test utterance.
For practical purposes, accumulated classification is more realistic than frame-wise classification.  Initially, a frame-wise assignment of noise classes is done, then assuming a constant noise source across the whole test signal, and then recursively dividing the test utterance into multiple segments, segment level classification is performed. It is assumed that at least 2 seconds of signal will have the same noise. The frame-wise class  label is assigned based on sum of cosine similarity for each dictionary using a block sparsity \cite{fu} based approach \cite{kv}. The steps for noise classification using this approach is enumerated in Algorithm \ref{recuralgo} .  A divide and conquer approach is proposed by recursively dividing the frames into two equal parts until 90\% of the component frames are classified as same class or the number of component frames corresponds to less than 2 seconds. All the segments are then assigned to two classes based on a rule based approach. In the case where 90\% of frames are not classified as same class, ten lowest energy frames within the segment are used to classify the segment. The reasoning behind using ten lowest energy frames is that in case the noise segment contains speech, the silence frames of speech (which have low energy) can be used for noise classification \cite{kv}. The exact transition instant is arrived at by refining the approximate transition frame.  Although the algorithm assumes that the signal consists of two consecutive noise segments, it  can be generalized to the segmentation of a signal containing multiple noises and transitions. The estimated noise classes are $\hat{m}_1,\;\hat{m}_2$ and the estimated transition frame is $i_t$.

\begin{algorithm}{Noise segmentation and classification}\label{recuralgo}

\begin{algorithmic}[1]

\State  $N_{ns}$ is the number of  noise source dictionaries, and $ c_{lab}=[]$ are  the estimated class labels initialized as null.
\State Find the frame-wise energy as $E(i)$ for $ 1\leq i\leq n$, for $n$ number of features.

\State Find the frame-wise sum of cosine-similarity of each feature $\mathbf{y}_i $ for each dictionary 
$\textbf{D}_j$ as
\begin{equation}
scs(i,j)=||\textbf{D}_j^T \mathbf{ y_i}||_1
\end{equation}

\State Sort the $ scs(i,j)$ in decreasing order in each row so as to get the initial estimate of noise indices  corresponding to the highest  cosine similarity as $mx_s$.

\State \Call{DivideRecursive}{$mx_s,1,n$}

\State Find the transition indices $tran$ where change of estimated noise  indices $ c_{lab}$ occurs
 \If{Number of elements of $tran > 1$ }
\State Assuming only two noise classes are present within the whole test signal, find two noise classes $\hat{m_1},\hat{m_2}$ which have the maximum occurence in $ c_{lab}$.
\State Find the centroid of the indices corresponding to the noise classes $\hat{m_1},\hat{m_2}$ in $ c_{lab}$ as $cent_{\hat{m}_1}, \;cent_{\hat{m}_2}$.
  \For{All the segments between transitions, $tran$}
\State Find the absolute differences between the centroid of the present segment, $cent_{seg}$  and $cent_{\hat{m}_1}, \;cent_{\hat{m}_2}$ as $dist_{\hat{m}_1}, \; dist_{ \hat{m}_2}$.

\State Update the class labels of the present segment as the index  $\hat{m}_1$ or $\hat{m}_2$  corresponding to minimum of $dist_{\hat{m}_1}, \; dist_{ \hat{m}_2}$
\EndFor
\EndIf

\State \textbf{Noise classification:} The estimated noise classes are $\hat{m_1},\hat{m_2}$, and initial estimate of transition frame is $i_{init}=index(change\; of\; c_{lab})$ 
\State $i_{init}$ is refined using the initial frame labels $mx_s$
\State \textbf{Noise segmentation:} Find the final transition frame within +/-1 second of $i_{init}$ around which there is equal distribution of  $\hat{m}_1,\hat{m}_2$ within a 2 second duration as  $i_t$.
\State The exact transition instant can be obtained from the mid-point of $ i_t$ frame in time domain.

\Function{DivideRecursive}{$mx_s ,a,b$}
\State Select a subset of the indices as $ind_{sel}=mx_s(a:b)$
 
\State Find the percentage occurrence of selected noise indices  $ind_{sel}$ among all the noise classes as  $ p_{occ}$
\State If the highest percentage occurrence  to a particular noise class $j$ is greater than $90\%$ i.e.
\If{$\max(p_{occ})\geq90$}
\State Update the estimate of the class labels as $c_{lab}(a:b)=j $
\Return
\ElsIf{ $(b-a)\leq nf_2$ ($nf_2$ is number of frames corresponding to 2 seconds)}

\State Sort the subset of frame energies $ E(a:b)$  and pick set of indices $ i_{10}$ corresponding to the ten lowest energy frames 
\cite{kv}
\State  Find the noise index $j$  having the highest occurrence among $ind_{sel}(i_{10})$
\State Update the estimate of the class labels as $c_{lab}(a:b)=j $
\Return

\EndIf

\State Assign $ m=\dfrac{a+b}{2}$
\State \Call{DivideRecursive}{$mx_s,a,m$}

\State \Call{DivideRecursive}{$mx_s,m+1,b$}
\EndFunction

\end{algorithmic}
\end{algorithm}

\subsection{Classification of speakers}

Given the estimated noise  sources, and the transition frame $ i_t$, the next task is to estimate the speaker corresponding to the utterance within each noise segment. The algorithm for speaker classification is explained in Algorithm \ref{spkalgo}.

The test feature $\textbf{y}$ is approximated  as the linear combination of the dictionary atoms from the estimated noise source, $\textbf{D}_{ns}^{\hat{m}_i}$ and the concatenated dictionary of speaker  sources $[\textbf{D}_{sp}^1.... \textbf{D}_{sp}^{N_{sp}}]$. Since speech segment occurs for a short duration and  consists of  silence, unvoiced and voiced regions, we use only those features having higher energy (30\% of the total number of features, $\textbf{y}$)  for speaker classification.  As speech is non-stationary and noise further corrupts it,  the speaker  index  is determined by comparing the weights estimated in the representation:

\begin{equation}\label{nsw}
\mathbf{y}\approx [\textbf{D}_{sp}^1.... \textbf{D}_{sp}^{N_{sp}} \textbf{D}_{ns}^{\hat{m}}][\mathbf{x}_1^T   ... \mathbf{x}_{N_{sp}}^T \mathbf{x}_{\hat{m}_i}^T]^T=\textbf{D}\mathbf{x}
\end{equation}
 The weight vector, $\textbf{x}$ is estimated by minimizing the distance $dist(\textbf{y},\textbf{Dx})$ using ASNA \cite{virtanen}, where $dist()$ is the KL-divergence between $\textbf{y}$ and $\textbf{Dx}$. 
  
   \begin{equation} 
  \stackunder{minimize}{\textbf{x}}\;
   KL(\mathbf{y}||\mathbf{\hat{y}}), \; \hat{\textbf{y}}=\textbf{D}\mathbf{x}  \;s.t.\;\mathbf{x}\geq 0 \label{asna}
   \end{equation}
  $\mathbf{x}$ is constrained to  be  non-negative and sparse by ASNA algorithm.
   
  A measure \textit{Sum of Weights (SW)} for each of the selected features $\mathbf{y}$ is defined as the absolute sum of elements of $\mathbf{x_k},\; 1\leq k \leq N_{sp}$ ,
  
  \begin{equation}\label{tsw}
  SW_k= || \mathbf{x}_k||_1
  \end{equation}
 
\begin{figure}[!h]
       \centering
      
       \includegraphics[width=.5\textwidth,height=.17\textheight]{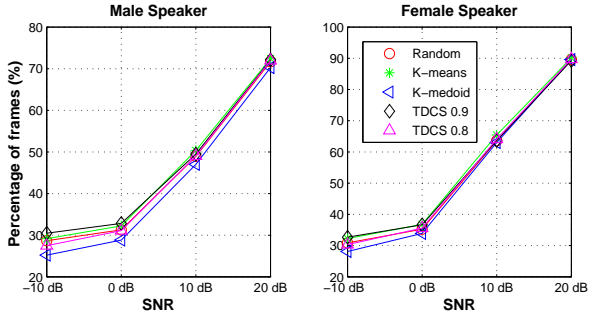}
       
       \caption{Percentage of 'high energy frames' employed  by the algorithm (chosen based on relative weights of speech and noise atoms)  for speaker classification, plotted as a function of input SNR in dB. 30 \% of estimated speech frames with energy higher than the rest are considered as high energy frames  }
       \label{percfr}
       \end{figure}

 The features used for speaker classification are further constrained by comparing the  sum of weights corresponding to the noise dictionary $SW_{\hat{m}_i}$ and the average $SW_k$ scaled by a factor, $fac=4$ corresponding to speaker dictionaries for each of the selected features. The reasoning is that the frames having approximate $SNR $ below -12 dB are neglected for speaker estimation. It is observed that we achieve an improvement of around 20\% for male and 10\% for female speaker classification accuracy using TDCS-0.8 dictionary method at 0 dB SNR by constraining the features used.    Figure \ref{percfr} shows the number of features constrained expressed  as a  percentage of selected frames as a function of SNR using different dictionary types for male and female speakers. It is observed that as SNR increases more number of features are used and the percentage is nearly  a linear function of SNR from 0 to 20 dB. So, lower number of features are desirable for speaker classification if the $SNR $ is low.  Also, it is seen that female speakers have higher percentage which means relative weights corresponding to speakers are higher for females.  The speaker sources are estimated as the indices $\hat{n}_1,\;\hat{n}_2$ as given in steps 9, 10 in Algorithm \ref{spkalgo}.

\begin{algorithm}{Speaker classification}\label{spkalgo}
\begin{algorithmic}[1]
\State Given the estimated noise sources $\hat{m_1},\;\hat{m_2}$ , divide the $n$ features into two sets $\mathbf{y}^1=\mathbf{y}_i\;s.t.\; 1\leq i\leq i_t$  and $\mathbf{y}^2=\mathbf{y}_i\;s.t.\; (1+i_t)\leq i\leq n$.
\State For each set of $\mathbf{y}^j,\;j\in[1,2]$, find the frame-wise energy  as $E(i)$.
\State Pick 30\% of features having highest energy, $E(i)$. 
\State Find the weights recovered using ASNA algorithm as in Equation \ref{nsw} for each of the features picked up in the previous step.
\State The  sum of weights, $SW_k$ corresponding to each dictionary index $k$ is found as in Equation \ref{tsw}.
\State Calculate the sum of weights corresponding to the noise dictionary as  $SW_{\hat{m}_i} $.
\State Find the set of frame indices, $ indst$  for which $SW_{\hat{m}_i} <fac\times \frac{\sum_k SW_k}{N_{sp}}, \;fac=4 $.  
\State In  case number of elements of   $ indst$  is very low, corresponding to less than around 700 ms, increment $fac$ until we get adequate $ indst$.

\State Find the total sum of weights using frame indices $ indst$ as   $TSW_k^i=\sum SW_k,  \forall\; \mathbf{y}=\mathbf{y}^j_i, \; i \in indst$.
\State Estimate the speaker index as $\hat{n}_i=\arg \max_k TSW_k^i$.

\end{algorithmic}
\end{algorithm}

\subsection{Speaker and noise dictionary update } \label{dicupd}

A novel algorithm to update speaker and noise dictionaries is proposed. It is a generalized algorithm which works with any dictionary learning algorithm. In the case of a test signal containing noise or speech only segments, the corresponding features  are used to update the estimated noise / speaker source dictionary. The intuition behind this method is that even though the dictionary for a particular source is not a good representation, it can be considered as the base dictionary which we update using  the test signal itself. It is also useful when the test signal is of short duration or it varies with time.     Algorithm \ref{dictalgo} illustrates the dictionary update algorithm in which a concatenation of the old dictionary and the test features $[\mathbf{D\; Y}]$ is used to update the dictionary as $ \mathbf{D}_{updat}$.

\begin{algorithm}
{Dictionary update}\label{dictalgo}
\begin{algorithmic}[1]
\item Given an input dictionary $\mathbf{D}$ and the test features  $\mathbf{Y}$
\item Update the dictionary $\mathbf{D}$ as $ \mathbf{D}_{u}=f([\mathbf{D\; Y}])$, where $f()$ is any dictionary learning algorithm like $K-means$

\end{algorithmic}
\end{algorithm}

\subsection{Separation of speech and noise signals}
The estimated noise indices $ \hat{m}_1,\;\hat{m}_2$ and speaker indices $ \hat{n}_1,\;\hat{n}_2$ are used to recover the noise and  speaker  components of the test features using a concatenated dictionary, $ \textbf{D}=[\textbf{D}_{ns}^{\hat{m}_i}\; \textbf{D}_{sp}^{\hat{n}_i} ]\forall \;i=1,2$ and recovery algorithm ASNA similar to eqn. \ref{asna}. The test features are divided into $\textbf{y}_1,\;\textbf{y}_2$ corresponding to the first and second parts using the estimated transition frame index $i_t$. The estimated features $\hat{y}_i$ and the noise and speech component $\hat{\textbf{y}}^i_{ns}, \; \hat{\textbf{y}}^i_{sp}$ are 

\begin{equation}
 \hat{\textbf{y}}_i=[\textbf{D}_{ns}^{\hat{m}_i}\; \textbf{D}_{sp}^{\hat{n}_i} ][\textbf{x}^T_{\hat{m}_i} \; \textbf{x}^T_{\hat{n}_i} ]^T
\end{equation}
\begin{equation}
\hat{\textbf{y}}^i_{ns}=\textbf{D}_{ns}^{\hat{m}_i}\textbf{x}^T_{\hat{m}_i} ,\;\hat{\textbf{y}}^i_{sp}=\textbf{D}_{sp}^{\hat{n}_i}\textbf{x}^T_{\hat{n}_i}
\end{equation}

The speech component in the time domain $s_{sp}[n]$ is reconstructed using the estimated $ \hat{\textbf{y}}^i_{sp}$ and phase of the mixed audio signal using  overlap and add method as $\hat{s}_{sp}[n]$ . The corresponding noise signal is estimated as $\hat{s}_{ns}=s_i[n]-\hat{s}_{sp}[n]$.

\subsubsection{Measures for quantifying separation performance }

The following measures are used to evaluate the performance of speech and noise separation:
\begin{itemize}
\item \textit{Signal to distortion ratio} (SDR) \cite{vincent} between the original and the estimated speech signal is defined as:

\begin{equation}
SDR=20\log_{10}\dfrac{||s_{sp}[n]||_2}{||s_{sp}[n]-\hat{s}_{sp}[n]||_2}
\end{equation}

SDR quantifies the deviation of the separated speech signal from the original signal; higher the SDR, better is the separation performance.
\item \textit{Error in segmental SNR}:  For segments in the  mixed signal, where speech is present, estimated segmental SNR is defined as the ratio of the total energy of the estimated speech to noise features in decibel, while the original segmental SNR uses the original features extracted from the ground truth speech and noise signal:

    \begin{eqnarray}
    SNR_{e}=10 \log  \dfrac{||\hat{\textbf{y}}_{sp}^i||^2_2}{||\hat{\textbf{y}}_{ns}^i||^2_2}\\    
    SNR_{o}=10\log\dfrac{||\textbf{y}_{sp}||^2_2}{||\textbf{y}_{ns}||^2_2 }
    \end{eqnarray}  
    
    The error in the estimate of SNR is given by $e_{SNR}=SNR_{o}- SNR_{e}$ . Mean of the absolute value of the  $e_{SNR}$ and the standard deviation of the  $e_{SNR}$ are used as the measures to quantify the performance of the algorithm.
 \end{itemize}
 
 It is to be noted that SDR and error in segmental SNR are computed only for the regions/ frames, where speech is present.
 \subsection{Detection of speech segments}

\begin{figure}[!h]
       \centering
      
       \includegraphics[width=.5\textwidth,height=.27\textheight]{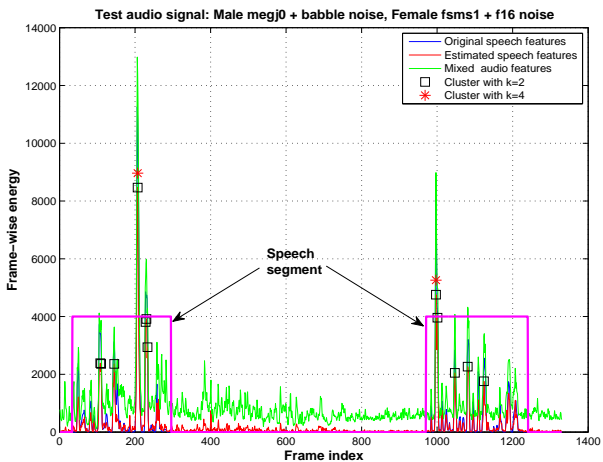}

       \caption{Illustration of detection of speech segments }
              \label{mrfar}
       \end{figure}

  Figure \ref{mrfar} shows the plot of  frame-wise energies for the  original, estimated and  mixed audio signal features and  speech segments in the mixed audio signal, at the SNR of 0 dB. The speech features are estimated using dictionaries of randomly  selected features of the   estimated speaker and  noise. It is seen from the figure that high energy frames  contain speech, since  voiced regions in the speech  have  high energy peaks.  In case the noise has almost non-varying energy, local maxima are high   in the speech segments. This distinguishes  the maxima in the speech and noise frames using a  k-means clustering algorithm to extract the significant maxima corresponding to the speech region .  
      The algorithm  used for extracting the speech segments is given  in Algorithm \ref{segalgo}  \cite{kv}.
     Two values of $k$, namely $2$ and $4$ are used as the number of clusters.   Figure \ref{mrfar} shows the local maxima and the clusters corresponding to the highest centroid as black squares and red stars. It can be seen that the cluster elements lie within the speech segments. It is to be noted that the clustering is carried out independently on the first and second parts   partitioned at the estimated noise transition frame of 732.
 
 \begin{algorithm}{Detection of the speech segments}\label{segalgo}
 \begin{algorithmic}[1]
 \State Extract local maxima of the frame energies.
 \State Do k-means clustering of the local maxima  using the initial  centroids as equally distributed between the maximum and minimum of the local maxima.
 \State Pick the  elements of the cluster corresponding to the highest centroid value and assign them as $cluster(k)$ which may correspond to the voiced segments of the speech.
 \end{algorithmic}
 \end{algorithm}

   Miss Rate and False alarm rate \cite{kv} are used as the measures for  the detection of the  speech segments:
    \begin{itemize}
    \item 
    Miss rate (MR): Percentage of number of speech segments, which do not encompass any element of $cluster(k)$ with respect to the total number of speech segments.
       \item False alarm rate (FAR): Percentage of  number  of $cluster(k)$ which are outside the speech segments with respect to the total number of $cluster(k)$.
     
    \end{itemize}

\section{Experimental setup}
This section describes the speaker and noise databases used and the training and test setup.
\subsection{Databases used}
For noise sources, all the 15 noises from the NOISEX database \cite{noisex} have been used. The first 20 seconds of each noise is used for testing and the rest for training. For speaker sources, ten male and ten female speakers are  randomly selected	 from dialect 5 of the training set of the TIMIT database. The first eight utterances from each speaker are used for training; the rest two are  used for  updating dictionary and  testing. Table \ref{dbase} shows the list of noise sources and male/ female speakers used in this work.

\subsection{Testing setup}
A mixed audio signal is simulated by concatenating two different noise sources such that the transition instant is between 9 and 11 seconds and the total duration is 20 seconds. A test utterance from a male speaker is added to the noise source in the first part, and from a female speaker to the second part at randomly selected locations  at SNR's of -10, 0, 10 and 20 dB.  
The minimum separation between speech utterances  is constrained to be 2 seconds. It is to be noted that our approach is independent of the gender of the speakers used in both parts of the test signal. Male and female speakers used in first and second part is only to get results on both male and female speakers separately.

Two different noise sources are randomly chosen for the first and second part such that all the noise sources are used in each part.  The speaker classes are selected such that all the ten  combinations are different and all male / female speakers are used. So, $15 \times10$ combinations of mixed audio signal are used for testing at different SNR's.

For testing our classification and separation performance using ground truth and updated dictionaries, we have four test cases using different combinations of dictionaries.
\begin{enumerate}
\item \textit{Complete speaker and noise dictionary :}  The test mixed audio signal is tested using all the noise and speaker dictionaries, and the separation is achieved using the identified noise/ speaker dictionary. 

\item \textit{Ground truth speaker and noise dictionary:} 
The ground truth speaker and noise dictionaries are used to evaluate the separation performance.

\item \textit{Out of set noise sources:}   The test mixed audio signal is tested using all the noise dictionaries except for the two dictionaries corresponding to the noise sources used in the test signal and all the speaker dictionaries. So, by pruning known noise dictionaries, the test signal is tested against out of set / unknown noise sources and known speakers. The results reported using this case show the robustness of our method given unknown test noises.
\item \textit{Out of set speaker sources:} The two dictionaries corresponding to the speaker sources used in the test audio signal are removed  from the  training set for classification and separation.  This case shows the robustness of our method given  unknown test speakers.

\end{enumerate}

 The out of set noise and speaker dictionaries used in the test cases $(3,\:4) $  give the estimated speaker and noise source indices. These dictionaries corresponding to the estimated source indices   are updated using the dictionary update method in Sec. \ref{dicupd}. For noise dictionary update, it is assumed that the features corresponding to noise only region is known and the same is used to update the estimated noise dictionary. For speaker dictionary update, features from the utterance not in the training / test set   is used to update the estimated speaker dictionary. Results on speaker classification  and separation performance are reported both before and after dictionary update using (estimated) out of set  and updated dictionaries.
    \begin{table}[!h]
    \centering
    \caption{Noise and speaker sources used }\resizebox{8cm}{!}{
\begin{tabular}{|c||c|c|}\hline
\textbf{Noise sources} & \multicolumn{2}{|c|}{\textbf{Speaker sources (TIMIT)}}\\
\hline
babble & \emph{\textbf{Male}} & \emph{\textbf{Female}}\\
\hline
buccaneer1 &megj0 &fsag0\\
\hline
buccaneer2 &mrkm0 &fmpg0\\
\hline
destroyerengine &mwch0 &fsdc0\\
\hline
destroyerops &mgsh0 &fsms1\\
\hline
f16 &mjpg0 &fdtd0\\
\hline
factory1 &mrew1 &fbjl0\\
\hline
factory2 &mfer0 &fcdr1\\
\hline
hfchannel &mwsh0 &fdmy0\\
\hline
leopard &mdhl0 &flmk0\\
\hline
m109 &mmcc0 &fbmh0\\
\hline
machinegun & &\\
\hline
pink & &\\
\hline
volvo & &\\
\hline
white & &\\
\hline
\end{tabular}
}
\label{dbase}\end{table}
\section{Results and discussion}
The results for noise and speaker classification, detection of noise transition instant, SDR, error in SNR, miss rate (MR) and false alarm rate (FAR)  are reported in this section. The five dictionary learning methods used are (1) random selection, (2) K-means, (3) K-medoid, (4) TDCS-0.9 and (5) TDCS-0.8. All the results given below are averages  over all the  combinations of  speakers and noises for male and female speakers separately, unless otherwise mentioned. It is to be noted that although we report separate results for male and female speakers, the test audio signal is tested against all twenty speaker dictionaries (both male and female) for speaker classification stage.

Table \ref{nscl} shows the overall noise classification accuracy at various $SNR's$ in the presence of utterances from male and female speakers separately. Random, K-means and TDCS-0.8 give the same accuracies for male and female speakers, so it is shown as a single value. All the noises are correctly classified except for 
machine gun noise which is mostly misclassified as volvo noise, which reduces the accuracy to 93.3\%. K-medoid gives the best accuracy for female speakers, while TDCS-0.9 gives the best accuracy for male speakers. In the case of TDCS-0.8, factory1 is also misclassified as pink noise, which reduces the accuracy to 86.67\%.

      \begin{table}[!h]
        \centering

        \caption{Noise classification accuracy using the five dictionary learning methods at  SNR of -10, 0, 10 and 20 dB } 
        
                \label{nscl} \resizebox{9cm}{!}{ 
        \begin{tabular}{|c|c|c|cc|cc|c|}
        \hline \textit{Dictionary} &  \textbf{Random}& \textbf{K-means} &  \multicolumn{2}{|c|}{\textbf{K-medoid}} &  \multicolumn{2}{|c|}{\textbf{TDCS 0.9 }} & \textbf{TDCS 0.8}\\
        & & & \textit{Male} & \textit{Female} & \textit{Male} & \textit{Female}& \\\hline
        -10 dB& 93.33 & 93.33 & 98.00 & 99.33 & 98.67 & 99.33 & 86.67 \\
        0 dB & 93.33 & 93.33 & 97.33 & 100.00 & 99.33 & 98.00 & 86.67 \\
        10 dB & 93.33 & 93.33 & 98.67 & 99.33 & 99.33 & 96.67 & 86.67 \\
        20 dB & 93.33 & 93.33 & 100.00 & 100.00& 100.00 & 98.67 & 86.67 \\
        \hline
        \end{tabular}}
        
        \end{table}

Table \ref{sca} shows the overall speaker classification accuracy using the complete noise and speaker dictionaries, out of set noise dictionaries,   updated noise dictionaries, and when the right speaker is within the top three speakers (based on  $TSW_k^i$ in Algorithm \ref{spkalgo}) at various SNR's  using the five dictionary learning methods. It is seen that  top three using complete gives the best accuracy across all the  cases. So, it helps in narrowing down to top three speakers, and other approaches can be used to classify the speaker among the top three.

    \begin{table}[!h]
    \centering
    \caption{Speaker classification accuracy using the complete noise / speaker dictionaries (Complete), out of set noise dictionaries (Unknown) and updated noise dictionary (Updated) and complete within top three speakers (Comp - Top 3) on male and female speakers for various dictionary methods at  SNR values of -10, 0, 10 and 20 dB}
    \label{sca}
  \subfloat[SNR= -10 dB]{%
   \resizebox{9cm}{!}{   \begin{tabular}{|c|cc|cc|cc|cc|cc|}\hline
\multirow{2}{*}{\textbf{Dictionary}}&     \multicolumn{2}{|c|}{  \textbf{Random} }&  \multicolumn{2}{|c|}{\textbf{K-means}} &  \multicolumn{2}{|c|}{\textbf{K-medoid}} &  \multicolumn{2}{|c|}{\textbf{TDCS 0.9}}& \multicolumn{2}{|c|}{\textbf{TDCS 0.8}}\\
&\emph{Male} &\emph{Female}&\emph{Male} &\emph{Female}&\emph{Male} &\emph{Female}&\emph{Male} &\emph{Female}&\emph{Male} &\emph{Female}\\\hline
\textbf{Complete}  & 25.33 & 29.33 & 27.33 & 27.33 & 21.33 & 22.00 & 26.67 & 18.00 & 22.00 & 18.67 \\
\textbf{Unknown} &16.00 & 12.67 & 14.67 & 12.67 & 14.00 & 11.33 & 14.00 & 8.00 & 13.33 & 10.67 \\
\textbf{Updated} & 29.33 & 30.00 & 29.33 & 29.33 & 19.33 & 21.33 & 27.33 & 18.00 & 23.33 & 22.00 \\
\textbf{Comp - Top 3} & 42.00 & 50.00 & 46.67 & 50.67 & 39.33 & 48.67 & 38.67 & 42.00 & 39.33 & 40.00 \\
\hline
\end{tabular}

}

    }

          \subfloat[SNR= 0 dB]{%
           \resizebox{9cm}{!}{   \begin{tabular}{|c|cc|cc|cc|cc|cc|}\hline
        \multirow{2}{*}{\textbf{Dictionary}}&     \multicolumn{2}{|c|}{  \textbf{Random} }&  \multicolumn{2}{|c|}{\textbf{K-means}} &  \multicolumn{2}{|c|}{\textbf{K-medoid}} &  \multicolumn{2}{|c|}{\textbf{TDCS 0.9}}& \multicolumn{2}{|c|}{\textbf{TDCS 0.8}}\\
        &\emph{Male} &\emph{Female}&\emph{Male} &\emph{Female}&\emph{Male} &\emph{Female}&\emph{Male} &\emph{Female}&\emph{Male} &\emph{Female}\\\hline

      \textbf{Complete}  &  67.33 & 63.33 & 72.00 & 58.00 & 57.33 & 48.67 & 57.33 & 26.00 & 74.67 & 65.33 \\
       \textbf{Unknown} & 43.33 & 49.33 & 42.67 & 44.67 & 38.67 & 36.67 & 36.00 & 23.33 & 45.33 & 35.33 \\
      \textbf{Updated} &   70.00 & 60.67 & 76.67 & 60.67 & 55.33 & 44.67 & 56.67 & 25.33 & 74.67 & 60.67 \\
      \textbf{Comp - Top 3} & 86.00 & 90.67 & 88.00 & 80.67 & 80.00 & 72.00 & 74.67 & 69.33 & 92.00 & 91.33 \\
        \hline
        \end{tabular}
        
        }}

             \subfloat[SNR= 10 dB]{%
                   \resizebox{9cm}{!}{   \begin{tabular}{|c|cc|cc|cc|cc|cc|}\hline
                \multirow{2}{*}{\textbf{Dictionary}}&     \multicolumn{2}{|c|}{  \textbf{Random} }&  \multicolumn{2}{|c|}{\textbf{K-means}} &  \multicolumn{2}{|c|}{\textbf{K-medoid}} &  \multicolumn{2}{|c|}{\textbf{TDCS 0.9}}& \multicolumn{2}{|c|}{\textbf{TDCS 0.8}}\\
                &\emph{Male} &\emph{Female}&\emph{Male} &\emph{Female}&\emph{Male} &\emph{Female}&\emph{Male} &\emph{Female}&\emph{Male} &\emph{Female}\\\hline

             \textbf{Complete}  &   86.67 & 78.67 & 98.00 & 75.33 & 78.67 & 70.00 & 79.33 & 34.00 & 96.00 & 72.67 \\
             \textbf{Unknown}  &   83.33 & 76.67 & 92.00 & 72.67 & 78.00 & 68.67 & 78.67 & 32.67 & 92.00 & 72.00 \\
             \textbf{Updated}  &  88.00 & 80.00 & 96.67 & 72.67 & 79.33 & 70.67 & 78.00 & 35.33 & 98.00 & 76.00 \\
              \textbf{Comp - Top 3}& 100.00 & 99.33 & 100.00 & 88.67 & 100.00 & 86.00 & 97.33 & 82.00 & 100.00 & 98.67 \\
                \hline
                \end{tabular}        
                }    
                }

      \subfloat[SNR= 20 dB]{%
       \resizebox{9cm}{!}{   \begin{tabular}{|c|cc|cc|cc|cc|cc|}\hline
    \multirow{2}{*}{\textbf{Dictionary}}&     \multicolumn{2}{|c|}{  \textbf{Random} }&  \multicolumn{2}{|c|}{\textbf{K-means}} &  \multicolumn{2}{|c|}{\textbf{K-medoid}} &  \multicolumn{2}{|c|}{\textbf{TDCS 0.9}}& \multicolumn{2}{|c|}{\textbf{TDCS 0.8}}\\
    &\emph{Male} &\emph{Female}&\emph{Male} &\emph{Female}&\emph{Male} &\emph{Female}&\emph{Male} &\emph{Female}&\emph{Male} &\emph{Female}\\\hline

   \textbf{Complete}  & 100.00 & 80.00 & 100.00 & 72.67 & 100.00 & 76.67 & 90.00 & 40.00 & 100.00 & 88.00 \\
    \textbf{Unknown}  & 100.00 & 80.00 & 100.00 & 72.67 & 100.00 & 76.67 & 90.00 & 41.33 & 100.00 & 86.00 \\
   \textbf{Updated}  &  100.00 & 80.00 & 100.00 & 74.67 & 100.00 & 78.00 & 90.00 & 40.67 & 100.00 & 87.33 \\
   \textbf{Comp - Top 3} & 100.00 & 100.00 & 100.00 & 92.00 & 100.00 & 89.33 & 100.00 & 85.33 & 100.00 & 100.00 \\
   
    \hline

    \end{tabular}
    
    }
    
        }
    
    \end{table}
    
        It is seen that random selection and K-means give the best speaker classification accuracy at SNR= -10 dB while other methods do not degrade much, relatively. TDCS-0.8 gives the best accuracy at all SNR's  due to the variability  of speech features requiring a low value threshold on cosine similarity measure, except for SNR= -10 dB . At an SNR of  20 dB, TDCS-0.9 gives low accuracy  for female speakers while lowering the threshold $T_w,T_b$  to 0.8 increases the accuracy to around 90\%. Unknown noise using out of set noise dictionaries gives lowest accuracies while using an updated dictionary gives similar accuracies as the complete dictionaries. It is seen that male speakers have  accuracy  higher than female speakers, and we get 100\% for male speakers at SNR= 20 dB. This may be due to the low pitch frequency of male speakers with the harmonics concentrated in the low frequency regions, which may not be corrupted with noise. 
        
         Noise classification accuracy is better at a threshold of 0.9 while speaker classification is better at 0.8. Thus, selection of appropriate  threshold is necessary.
         
         Table \ref{err} shows the mean absolute and standard deviation of error (in seconds) in the detection of noise transition instant for various SNR's. It is seen than TDCS-0.9 gives the lowest mean absolute and standard deviation of error of 0.16 seconds at SNR= -10 dB, while there is a slight increase of error with increase in SNR.

              \begin{table}[!h]
                \centering

                \caption{Mean absolute error and standard deviation of error (in seconds) in the detection of  noise transition instant, for various SNR's and dictionary learning methods } 
                \label{err} \resizebox{9cm}{!}{ 
                \begin{tabular}{|c|cc|cc|cc|cc|cc|}
                \hline
                  \multirow{2}{*}{\textbf{Dictionary}}&     \multicolumn{2}{|c|}{  \textbf{Random} }&  \multicolumn{2}{|c|}{\textbf{K-means}} &  \multicolumn{2}{|c|}{\textbf{K-medoid}} &  \multicolumn{2}{|c|}{\textbf{TDCS 0.9}}& \multicolumn{2}{|c|}{\textbf{TDCS 0.8}}\\
                   & \emph{Mean } &\emph{STD}&\emph{Mean } &\emph{STD}&\emph{Mean} &\emph{STD}&\emph{Mean} &\emph{STD}&\emph{Mean} &\emph{STD}\\
                \hline
               -10 dB& 0.20 & 0.28 & 0.21 & 0.34 & 0.25 & 0.42 & 0.16 & 0.24 & 0.22 & 0.28 \\
                0 dB & 0.20 & 0.29 & 0.22 & 0.34 & 0.25 & 0.42 & 0.17 & 0.26 & 0.23 & 0.28 \\
                10 dB & 0.21 & 0.29 & 0.22 & 0.34 & 0.26 & 0.43 & 0.19 & 0.28 & 0.24 & 0.30 \\
                20 dB & 0.21 & 0.29 & 0.22 & 0.35 & 0.25 & 0.42 & 0.18 & 0.27 & 0.23 & 0.29 \\
                \hline
                \end{tabular}
                
                }                
                \end{table}

\begin{figure}[!h]
      \centering
     
      \includegraphics[width=.5\textwidth,height=.35\textheight]{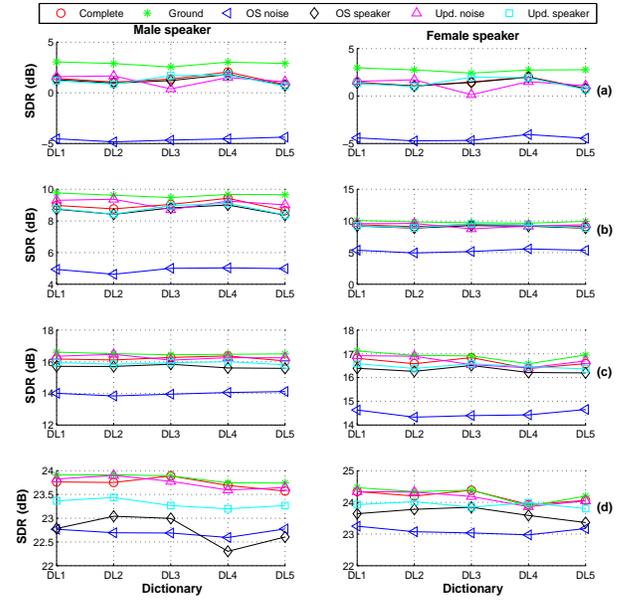}
      
      \caption{Plots of  SDR for input  SNR's of (a) -10, (b) 0, (c) 10 and (d) 20 dB using test cases  as complete dictionaries (Complete), ground truth dictionaries (Ground), out of set noise (OS noise) , out of set speaker (OS speaker), updated noise dictionary (Upd. noise) and updated speaker dictionary (Upd. speaker). Dictionary learning methods are denoted by DL1: Random, DL2: K-means, DL3: K-medoid, DL4: TDCS-0.9, DL5: TDCS-0.8.}
      \label{sdrpl}
      \end{figure}

Figure \ref{sdrpl} shows the variation of SDR in dB using the five dictionary  methods evaluated on complete dictionaries (Complete), ground truth dictionaries (Ground), out of set noise (OS noise), out of set speaker dictionaries  (OS speaker), updated noise (Upd. noise) and updated speaker dictionary (Upd. speaker). It is seen that using Ground givesthe  best SDR while  using OS noise dictionaries gives the lowest SDR. Upd. noise dictionary gives SDR comparable to Complete test cases. Using OS speaker results in the lowest SDR at a SNR of 20 dB due to high speech energy. It is observed that using OS speaker does not degrade SDR much as compared to Complete at other SNR's. Average improvements of SDR over SNR of -10, 0, 10 and 20 dB are 11, 9, 6 and 3 dB, neglecting the worst case using OS noise. The variation of SDR is not much across the different dictionary methods.

\begin{figure}[!h]
            \centering
           
            \includegraphics[width=.5\textwidth,height=.35\textheight]{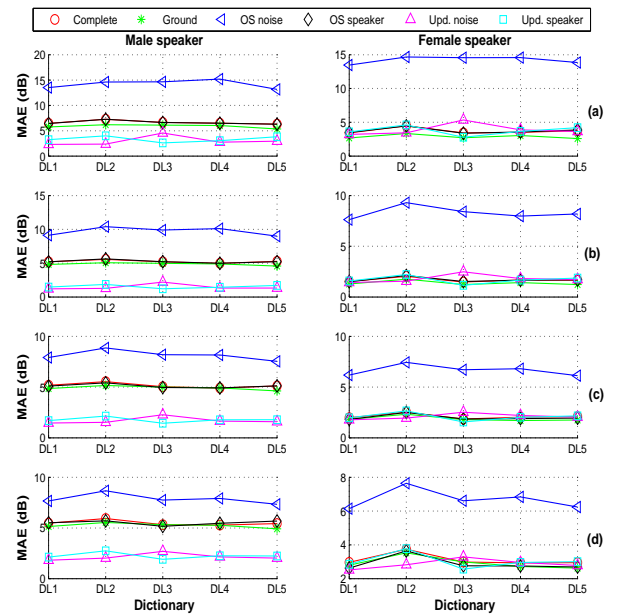}
            
            \caption{Plots of  mean absolute error (MAE) between the original and the estimated segmental SNR  for input  SNR of (a) -10, (b) 0 , (c) 10 and (d) 20 dB. Notations are the same as in Fig. \ref{sdrpl}. }
            \label{errsnrmn}
            \end{figure}

\begin{figure}[!h]
                        \centering
                       
                        \includegraphics[width=.5\textwidth,height=.35\textheight]{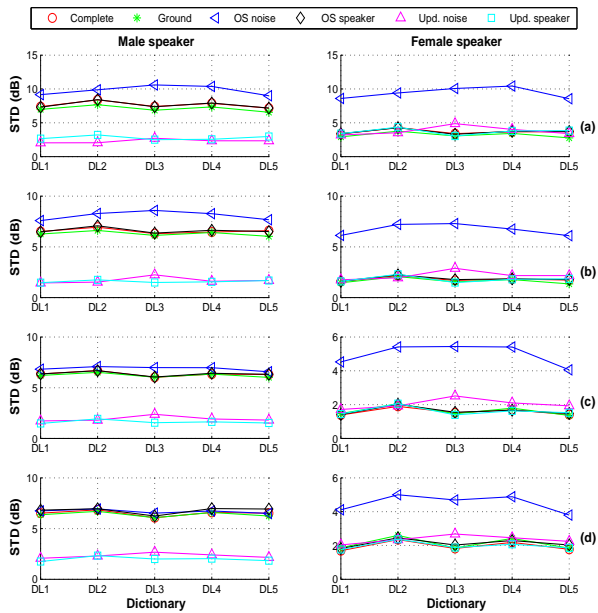}
                        
                        \caption{Illustration of  standard deviation of error (STD) between the original and the estimated segmental SNR. Notations are the same as in Fig. \ref{sdrpl}. }
                        \label{errsnrstd}
                        \end{figure}

Figure \ref{errsnrmn}  shows the variation of mean absolute error (MAE)  while  Fig. \ref{errsnrstd} shows the standard deviation of error (STD) for different dictionary learning methods and test cases similar to Fig. \ref{sdrpl}. It is seen that for male speakers, MAE and STD for updated noise and updated speaker dictionary cases outperform other cases at all SNR's for male speakers. OS noise results in the worst performance over all test cases giving an  MAE of around 15 dB at -10 dB. Also, we get MAE and STD  of around 2 dB as the best over all the test cases.

      Table \ref{mr} shows the miss rate (MR) while Table \ref{far}  shows the false alarm rate (FAR) in percentage for the five dictionary methods and four test cases for the choices of number of clusters $k=2$ and $4$. It is observed complete and ground truth dictionaries achieve poor performance at SNR of -10 dB even though these test cases give better results on other measures. It may be due to the few high noise energy segments at low SNR's. At high SNR's, it is seen that MR is almost negligible and zero for 10, 20 dB SNR except for OS noise test case. It is seen that high FAR is obtained at -10 dB SNR due to the high variability in the energy of some of the noises like machine gun. Using $k=4$ improves the FAR by 10-20 \%  while it degrades MR by around 2- 15\%. FAR is zero at high SNR's while it is less than about 10\% at 0 dB SNR. FAR is the highest for  OS noise test case at a SNR of -10 dB.
      
      \subsection{Comparison with previous work}
      The results presented are not directly comparable to other methods in the literature since the test cases simulated in this paper are unique and novel. A few of the results are compared indirectly here. Joder et. al. \cite{joder}  reported a speaker classification accuracy of 98.9 \% for eight speakers with clean speech, while we achieve 100\% accuracy for male speakers and 88\% for female speakers using TDCS-0.8 method at  SNR= 20 dB SNR. Rose et. al. \cite{rose} proposed speaker identification in noise using Gausssian mixture models for 16 speakers and 10 seconds segments reporting a accuracy of 79.9 \% at 10 dB SNR  while we get average accuracy of 80\% tested against 20 speakers at 10 dB in the presence of babble noise. On white noise, \cite{rose} reported 68.8\% accuracy while we get 90\% accuracy at 10 dB SNR. So, we achieve a comparably higher speaker classification accuracy even though we do not know the location of the speech segments.  Loizou \cite{loizou} performed speech enhancement and reported an improvement in segmental SNR    in speech with speech-shaped noise of around 5 dB at 0 dB SNR while we achieve a high  SDR of around 10 dB at 0 dB SNR, which is equivalent an to improvement of 10 dB.       
       Mohammadiha et.al. \cite{mohammadiha} did unsupervised speech enhancement based on Bayesian formulation of NMF  reported SDR of around 5.5 dB at 0 dB  SNR while we reported SDR of around 9 dB using our methods in all test cases except for OS noise case.

          \begin{table}[!h]
          \centering
          \caption{Miss rate (MR) for the detection of speech using different dictionary methods on Complete, Ground, OS noise and OS speaker test cases. Only OS noise test case is shown for 10 and 20 dB as MR  is zero for all other  cases.}
          \label{mr}
        \subfloat[SNR= -10 dB]{%
         \resizebox{9cm}{!}{   \begin{tabular}{|cc|cc|cc|cc|cc|cc|}\hline
     \multicolumn{2}{|c|}{   \multirow{2}{*}{\textbf{Dictionary}}}&     \multicolumn{2}{|c|}{  \textbf{Random} }&  \multicolumn{2}{|c|}{\textbf{K-means}} &  \multicolumn{2}{|c|}{\textbf{K-medoid}} &  \multicolumn{2}{|c|}{\textbf{TDCS 0.9}}& \multicolumn{2}{|c|}{\textbf{TDCS 0.8}}\\
    &  &\emph{Male} &\emph{Female}&\emph{Male} &\emph{Female}&\emph{Male} &\emph{Female}&\emph{Male} &\emph{Female}&\emph{Male} &\emph{Female}\\\hline
      
   \multirow{2}{*}{\textbf{Complete}}  & \textbf{k=2 }&   12.00 & 1.33 & 12.00 & 0.00 & 12.00 & 3.33 & 9.33 & 3.33 & 11.33 & 1.33 \\
  &\textbf{k=4}&    14.67 & 5.33 & 15.33 & 8.00 & 19.33 & 18.00 & 20.00 & 8.00 & 17.33 & 8.67 \\ \hline
 \multirow{2}{*}{\textbf{Ground}}  & \textbf{k=2 }   &  10.67 & 2.00 & 10.00 & 2.00 & 12.67 & 4.00 & 12.00 & 2.00 & 10.67 & 2.67 \\
     &\textbf{k=4} & 14.67 & 8.67 & 13.33 & 9.33 & 16.67 & 16.00 & 20.67 & 6.67 & 13.33 & 5.33 \\ \hline
 \multirow{2}{*}{\textbf{OS noise}}  & \textbf{k=2 }   &  0.00 & 0.00 & 2.00 & 0.00 & 2.67 & 3.33 & 3.33 & 3.33 & 3.33 & 4.67 \\
    &\textbf{k=4} & 6.67 & 9.33 & 6.00 & 11.33 & 9.33 & 10.67 & 8.67 & 14.00 & 11.33 & 16.00 \\\hline
  \multirow{2}{*}{\textbf{OS speaker}}  & \textbf{k=2 }   & 10.67 & 0.67 & 9.33 & 0.00 & 11.33 & 4.00 & 11.33 & 2.67 & 9.33 & 2.67 \\
     &\textbf{k=4}&   14.00 & 6.67 & 14.00 & 8.00 & 21.33 & 15.33 & 20.00 & 8.00 & 16.67 & 8.67 \\\hline

      \end{tabular}}}

          \subfloat[SNR= 0 dB]{%
               \resizebox{9cm}{!}{   \begin{tabular}{|cc|cc|cc|cc|cc|cc|}\hline
           \multicolumn{2}{|c|}{   \multirow{2}{*}{\textbf{Dictionary}}}&     \multicolumn{2}{|c|}{  \textbf{Random} }&  \multicolumn{2}{|c|}{\textbf{K-means}} &  \multicolumn{2}{|c|}{\textbf{K-medoid}} &  \multicolumn{2}{|c|}{\textbf{TDCS 0.9}}& \multicolumn{2}{|c|}{\textbf{TDCS 0.8}}\\
          &  &\emph{Male} &\emph{Female}&\emph{Male} &\emph{Female}&\emph{Male} &\emph{Female}&\emph{Male} &\emph{Female}&\emph{Male} &\emph{Female}\\\hline
            
         \multirow{2}{*}{\textbf{Complete}}  & \textbf{k=2 }&  0.00 & 0.00 & 0.00 & 0.00 & 0.00 & 0.00 & 0.00 & 0.00 & 0.00 & 0.00  \\
        &\textbf{k=4}&   0.00 & 0.00 & 0.00 & 0.00 & 0.00 & 0.00 & 0.00 & 0.00 & 1.33 & 0.00  \\ \hline
       \multirow{2}{*}{\textbf{Ground}}  & \textbf{k=2 }   & 0.00 & 0.00 & 0.00 & 0.00 & 0.00 & 0.00 & 0.00 & 0.00 & 0.00 & 0.00 \\ 
           &\textbf{k=4} & 0.00 & 0.00 & 0.00 & 0.00 & 0.67 & 0.00 & 0.00 & 0.00 & 0.00 & 0.00  \\ \hline
       \multirow{2}{*}{\textbf{OS noise}}  & \textbf{k=2 }   & 0.00 & 0.00 & 0.00 & 0.00 & 0.00 & 0.00 & 0.00 & 0.67 & 0.00 & 0.67  \\
          &\textbf{k=4} & 0.00 & 2.00 & 0.00 & 1.33 & 0.00 & 1.33 & 0.00 & 2.00 & 0.00 & 3.33  \\\hline
        \multirow{2}{*}{\textbf{OS speaker}}  & \textbf{k=2 }   & 0.00 & 0.00 & 0.00 & 0.00 & 0.00 & 0.00 & 0.00 & 0.00 & 0.00 & 0.00  \\
           &\textbf{k=4}&  0.00 & 0.00 & 0.67 & 0.00 & 0.67 & 0.00 & 0.00 & 0.00 & 0.67 & 0.00 \\\hline

            \end{tabular}}}

              \subfloat[SNR= 10 dB]{%
                           \resizebox{9cm}{!}{   \begin{tabular}{|cc|cc|cc|cc|cc|cc|}\hline
                       \multicolumn{2}{|c|}{   \multirow{2}{*}{\textbf{Dictionary}}}&     \multicolumn{2}{|c|}{  \textbf{Random} }&  \multicolumn{2}{|c|}{\textbf{K-means}} &  \multicolumn{2}{|c|}{\textbf{K-medoid}} &  \multicolumn{2}{|c|}{\textbf{TDCS 0.9}}& \multicolumn{2}{|c|}{\textbf{TDCS 0.8}}\\
                      &  &\emph{Male} &\emph{Female}&\emph{Male} &\emph{Female}&\emph{Male} &\emph{Female}&\emph{Male} &\emph{Female}&\emph{Male} &\emph{Female}\\\hline
                        
                     \multirow{2}{*}{\textbf{OS noise}}  & \textbf{k=2 }& 0.00 & 0.00 & 0.00 & 0.00 & 0.00 & 0.00 & 0.00 & 0.67 & 0.00 & 0.00 \\
                    &\textbf{k=4}&  0.00 & 2.00 & 0.00 & 1.33 & 0.00 & 1.33 & 0.00 & 1.33 & 0.00 & 2.67  \\ \hline
                
         \end{tabular}}}
                  
                   \subfloat[SNR= 20 dB]{%
                                          \resizebox{9cm}{!}{   \begin{tabular}{|cc|cc|cc|cc|cc|cc|}\hline
                                      \multicolumn{2}{|c|}{   \multirow{2}{*}{\textbf{Dictionary}}}&     \multicolumn{2}{|c|}{  \textbf{Random} }&  \multicolumn{2}{|c|}{\textbf{K-means}} &  \multicolumn{2}{|c|}{\textbf{K-medoid}} &  \multicolumn{2}{|c|}{\textbf{TDCS 0.9}}& \multicolumn{2}{|c|}{\textbf{TDCS 0.8}}\\
                                     &  &\emph{Male} &\emph{Female}&\emph{Male} &\emph{Female}&\emph{Male} &\emph{Female}&\emph{Male} &\emph{Female}&\emph{Male} &\emph{Female}\\\hline
                                       
                                    \multirow{2}{*}{\textbf{OS noise}}  & \textbf{k=2 }& 0.00 & 0.00 & 0.00 & 0.00 & 0.00 & 0.00 & 0.00 & 0.00 & 0.00 & 0.00  \\
                                   &\textbf{k=4}&  0.00 & 2.67 & 0.00 & 1.33 & 0.00 & 1.33 & 0.00 & 2.00 & 0.00 & 2.67  \\ \hline
                               
                        \end{tabular}}}
        \end{table}

                \begin{table}[!h]
                \centering
                \caption{False alarm rate (FAR) for the detection of speech segments  using different dictionary methods on Complete, Ground, OS noise and OS speaker test cases. Only OS noise test case is shown for 10 dB, and no cases for 20 dB as FAR  is zero for all other cases. }
                \label{far}
                    \subfloat[SNR= -10 dB]{%
                               \resizebox{9cm}{!}{   \begin{tabular}{|cc|cc|cc|cc|cc|cc|}\hline
                           \multicolumn{2}{|c|}{   \multirow{2}{*}{\textbf{Dictionary}}}&     \multicolumn{2}{|c|}{  \textbf{Random} }&  \multicolumn{2}{|c|}{\textbf{K-means}} &  \multicolumn{2}{|c|}{\textbf{K-medoid}} &  \multicolumn{2}{|c|}{\textbf{TDCS 0.9}}& \multicolumn{2}{|c|}{\textbf{TDCS 0.8}}\\
                          &  &\emph{Male} &\emph{Female}&\emph{Male} &\emph{Female}&\emph{Male} &\emph{Female}&\emph{Male} &\emph{Female}&\emph{Male} &\emph{Female}\\\hline
                            
                         \multirow{2}{*}{\textbf{Complete}}  & \textbf{k=2 }& 48.99 & 33.36 & 50.94 & 36.23 & 50.78 & 33.77 & 41.24 & 31.16 & 49.21 & 39.58  \\
                        &\textbf{k=4}&   27.42 & 24.10 & 31.19 & 20.96 & 41.67 & 28.57 & 35.94 & 17.73 & 36.87 & 29.75\\\hline
                       \multirow{2}{*}{\textbf{Ground}}  & \textbf{k=2 }  & 35.28 & 25.60 & 40.89 & 24.40 & 46.20 & 28.61 & 37.15 & 27.34 & 33.67 & 24.13 \\
                           &\textbf{k=4} & 25.00 & 15.33 & 23.08 & 14.51 & 32.75 & 20.85 & 27.68 & 14.67 & 19.92 & 15.56  \\\hline
                       \multirow{2}{*}{\textbf{OS noise}}  & \textbf{k=2 }   & 53.78 & 49.45 & 53.85 & 48.72 & 55.95 & 49.75 & 53.80 & 47.91 & 53.88 & 49.77 \\
                          &\textbf{k=4} & 36.15  & 30.97 & 35.62 & 34.74 & 36.50 & 36.88 & 34.55 & 34.54 & 34.99 & 30.97 \\\hline
                        \multirow{2}{*}{\textbf{OS speaker}}  & \textbf{k=2 }   & 47.60 & 33.02 & 49.45 & 36.97 & 49.70 & 33.09 & 40.79 & 31.93 & 49.62 & 39.39 \\
                           &\textbf{k=4}&  28.25 & 24.63 & 30.12 & 20.85 & 40.38 & 29.10 & 34.72 & 18.61 & 36.90 & 27.66 \\\hline

                            \end{tabular}}}

              \subfloat[SNR= 0 dB]{%
               \resizebox{9cm}{!}{   \begin{tabular}{|cc|cc|cc|cc|cc|cc|}\hline
           \multicolumn{2}{|c|}{   \multirow{2}{*}{\textbf{Dictionary}}}&     \multicolumn{2}{|c|}{  \textbf{Random} }&  \multicolumn{2}{|c|}{\textbf{K-means}} &  \multicolumn{2}{|c|}{\textbf{K-medoid}} &  \multicolumn{2}{|c|}{\textbf{TDCS 0.9}}& \multicolumn{2}{|c|}{\textbf{TDCS 0.8}}\\
          &  &\emph{Male} &\emph{Female}&\emph{Male} &\emph{Female}&\emph{Male} &\emph{Female}&\emph{Male} &\emph{Female}&\emph{Male} &\emph{Female}\\\hline
            
         \multirow{2}{*}{\textbf{Complete}}  & \textbf{k=2 }&  7.85 & 4.66 & 9.76 & 4.33 & 6.02 & 0.45 & 4.46 & 0.00 & 10.99 & 6.43 \\
        &\textbf{k=4}&  1.11 & 0.00 & 2.15 & 1.08 & 1.71 & 0.00 & 1.14 & 0.00 & 2.16 & 0.37 \\\hline
       \multirow{2}{*}{\textbf{Ground}}  & \textbf{k=2 }   &  5.00 & 0.36 & 5.24 & 0.09 & 5.15 & 0.52 & 6.00 & 0.26 & 5.13 & 0.09 \\
           &\textbf{k=4} &1.62 & 0.00 & 2.05 & 0.00 & 2.20 & 0.00 & 1.67 & 0.00 & 1.63 & 0.00  \\\hline
       \multirow{2}{*}{\textbf{OS noise}}  & \textbf{k=2 }   & 0.17 & 2.91 & 0.57 & 3.86 & 0.39 & 3.77 & 0.19 & 1.81 & 0.53 & 6.83 \\
          &\textbf{k=4} & 0.00 & 3.65 & 0.00 & 1.82 & 0.00 & 4.42 & 0.00 & 1.49 & 0.00 & 4.48  \\\hline
        \multirow{2}{*}{\textbf{OS speaker}}  & \textbf{k=2 }   & 9.23 & 6.90 & 11.67 & 6.98 & 5.28 & 0.10 & 7.07 & 0.00 & 13.94 & 8.01 \\
           &\textbf{k=4}&  3.07 & 0.79 & 1.20 & 0.00 & 1.70 & 0.00 & 1.21 & 0.00 & 4.32 & 0.39 \\\hline

            \end{tabular}}}
            
                     \subfloat[SNR= 10 dB]{%
                           \resizebox{9cm}{!}{   \begin{tabular}{|cc|cc|cc|cc|cc|cc|}\hline
                       \multicolumn{2}{|c|}{   \multirow{2}{*}{\textbf{Dictionary}}}&     \multicolumn{2}{|c|}{  \textbf{Random} }&  \multicolumn{2}{|c|}{\textbf{K-means}} &  \multicolumn{2}{|c|}{\textbf{K-medoid}} &  \multicolumn{2}{|c|}{\textbf{TDCS 0.9}}& \multicolumn{2}{|c|}{\textbf{TDCS 0.8}}\\
                      &  &\emph{Male} &\emph{Female}&\emph{Male} &\emph{Female}&\emph{Male} &\emph{Female}&\emph{Male} &\emph{Female}&\emph{Male} &\emph{Female}\\\hline
                        
                     \multirow{2}{*}{\textbf{OS noise}}  & \textbf{k=2 }&  0.00 & 2.98 & 0.00 & 2.41 & 0.00 & 3.24 & 0.00 & 2.16 & 0.00 & 3.10 \\
                    &\textbf{k=4}&   0.00 & 3.23 & 0.00 & 1.87 & 0.00 & 1.84 & 0.00 & 1.13 & 0.00 & 4.09  \\\hline
                
                            \end{tabular}}}
      \end{table}

\section{Conclusion}
A novel approach is proposed for the classification and separation of mixed audio signals commonly occurring in telephonic conversations. Since mobile communication is prolific nowadays, our approach can be used for tracking of speaker/ noise sources and noise adaptive speech enhancement using sparse representation based methods. We have shown how updation of dictionaries using parts of the test signal itself improves the classification and separation performance. As a future work, we plan to use machine learning techniques to learn discriminative dictionaries so as to classify multiple classes of noise and speech signals, and mixed audio signals. Using discriminative dictionaries may classify the various components in a mixed signal like language, speaker, gender, music and noises in a more generic way.


%


\ifCLASSOPTIONcaptionsoff
  \newpage
\fi


\begin{thebibliography}{9}
    \bibitem{turner} R Turner, [Online] \textit{http://www.wired.co.uk/news/archive/2013-10/02/machine-hearing-cambridge-university}
       
             
     
                        
                              \bibitem{ikram} S. Ikram, ``
                                     Digital audio forensics using background noise," \emph{Multimedia and Expo (ICME)}, July 2010, pp. 106-110.
                                     
           
               \bibitem{chu1}  S. Chu, S. Narayanan, C. C. Jay Kuo, and M. J. Matari, ``Where am I? Scene recognition for mobile robots using audio
                features," \emph{In IEEE International Conference on Multimedia and Expo}, pp. 885-888, 2006.
               
                \bibitem{eldar} Y. C. Eldar, P. Kuppinger and H. Bolcskei, ``Block-Sparse Signals: Uncertainty Relations and Efficient Recovery," \emph{IEEE Trans. Signal. Process.}, vol. 58, no-6, pp. 3042 - 3054, 2010. 
                
                
                           \bibitem{virtanen} T. Virtanen, J. F. Gemmeke, B. Raj, ``Active-set Newton algorithm for overcomplete non-negative representations of audio", \emph{IEEE Trans. Audio, Speech, and Lang. Process.}, vol. 21, pp. 2277 - 2289, 2013.
            
    
        \bibitem{lu} L. Lu and  H.Jiang, ``Content Analysis for Audio Classification and Segmentation," \textit{IEEE Trans. Speech Audio Processing}, vol. 10, no. 7,  2002.
        
   \bibitem{zhang} T. Zhang and  C. C. J. Kuo, ``Audio Content Analysis for Online Audiovisual Data Segmentation and Classification," \textit{IEEE Trans. Speech Audio Processing}, vol. 9, no. 4,  2001.
     
    
    \bibitem{barchesi}
            D. Barchiesi,  D. Giannoulis, D. Stowell, M. D. Plumbley and P. Mermelstein,
                   ``Acoustic Scene Classification: Classifying environments from the sounds they produce,''   \textit{IEEE Signal Processing Magazine}, vol.~32, no.~3, pp.~16--34,  2015.
    
     
                \bibitem{giannoulis} D. Giannoulis, E. Benetos, D. Stowell, M. Rossignol, M. Lagrange and M. D. Plumbley, ``Detection and classification of acoustic scenes and events: an IEEE AASP challenge," \emph{IEEE Workshop Applications of Signal Processing to Audio and Acoustics}, Oct. 2013.
                 
                    \bibitem{cauchi} B. Cauchi, ``Non-negative matrix factorization applied to auditory scene classification," \emph{ Master’s thesis, ATIAM (UPMC / IRCAM / TELECOM ParisTech)}, 2011.
                            
                            
         
                    
                    \bibitem{lyon} R. H. Lyon, ``Machinery Noise and Diagnostics,"  \emph{Butterworth-Heinemann}, 1987.
             
                    
                    \bibitem{shirkhodaie} A. Shirkhodaie, and A. Alkilani, ``A survey on acoustic signature recognition and classification techniques for persistent surveillance systems," \emph{Proc. Signal Processing, Sensor Fusion, and Target Recognition}, May 2012.
                    
                    
                    \bibitem{kates}
                    J. M. Kates, ``Classification of background noises for hearing aid applications," \emph{J. Acoust. Soc. Am.}, vol. 91 (1), pp. 461-470, Jan. 1995.
                    
                       \bibitem{maleh} K. El-Maleh, A. Samouelian, and P. Kabal, ``Frame-level noise classification in mobile environments,"  \emph{Proc. IEEE Conf. Acoustics, Speech, Signal Proc}, March 1999, pp. 237-240.
                    
                    \bibitem{casey} M. Casey, ``Reduced-rank spectra and minimum-entropy
                    priors as consistent and reliable cues for generalized sound
                    recognition," 	\emph{Proc. Workshop on Consistent and Reliable
                    Acoustic Cues for Sound Analysis, Eurospeech}, Aalborg,
                    Denmark, 2001.
                       
                    \bibitem{chu} S. Chu, S. Narayanan  and C. C. J. Kuo, ``Environmental sound recognition with time-frequency audio features," \emph{IEEE Trans. Audio, Speech, and Lang. Process.}, vol.17, no.6, 2009.
                    
                        \bibitem{chachada} S. Chachada  and C. C. J. Kuo, ``
                         Environmental sound recognition: A survey," \textit{Signal and Information Processing Association Annual Summit and Conference (APSIPA)}, 2013, p. 1--9.
                         
                         \bibitem{malik} H. Malik, ``Acoustic Environment Identiﬁcation and Its Applications to Audio Forensics," \textit{IEEE Trans. Information Forensics and Security}, vol. 8, no. 11, 2013
                          
         \bibitem{tzagkarakis} C. Tzagkarakis and A. Mouchtaris, ``Sparsity based robust speaker identification using a discriminative dictionary learning approach," \textit{Signal Processing Conference (EUSIPCO)}, 2013, pp. 1--5.
         
         
         \bibitem{joder} C. Joder and  B. Schuller ``Exploring Nonnegative Matrix Factorization for Audio Classiﬁcation: Application to Speaker Recognition," \textit{Proceedings of Speech Communication}, 2012, pp. 1--4.
         
            \bibitem{malkin} R. G. Malkin, ``Machine listening for context-aware 	computing," \emph{Doctoral Dissertation, Carnegie Mellon University}, 2006
%
%
%
%
%
%
                                  
                               \bibitem{virtanen0} T. Virtanen, ``Monaural sound source separation by non-negative matrix
                                 factorization with temporal continuity and sparseness criteria," \emph{IEEE Trans. Audio, Speech, and Lang. Process.}, vol.15, no.3, 2007.
                                 
                                 \bibitem{ozerov} A. Ozerov, ``Multichannel nonnegative matrix factorization in convolutive mixtures for audio source separation," \emph{IEEE Trans. Audio, Speech, and Lang. Process.}, vol. 18, no. 3, 2010.
                                 
                                 \bibitem{mysore}  G. J. Mysore,  P. Smaragdis, and  B. Raj, `` Non-negative Hidden Markov Modeling of Audio with Application to Source Separation," \emph{Lecture Notes in Computer Science,  Latent Variable Analysis and Signal Separation}, vol. 7572,  pp. 186-199, 2012. 
                                 
                                 \bibitem{bertin} N. Bertin, R. Badeau, and E. Vincent, ``Enforcing harmonicity and
                                   smoothness in Bayesian non-negative matrix factorization applied to
                                   polyphonic music transcription," \emph{IEEE Trans. Audio, Speech, and Lang. Process.}, vol. 18, no. 3, 2010.
                                   
                                 \bibitem{gemmeke} J. Gemmeke, T. Virtanen, and A. Hurmalainen, ``Exemplar-based sparse
                                   representations for noise robust automatic speech recognition," \emph{IEEE Trans. Audio, Speech, and Lang. Process.}, vol. 19, no. 7, 2011.
                                 
                                   \bibitem{raj} B. Raj, T. Virtanen, S. Chaudhuri, and R. Singh, ``Non-negative matrix
                                   factorization based compensation of music for automatic speech recognition," \emph{Interspeech 2010}, Tokyo, Japan, 2010.
                                 
                                 \bibitem{cho} Y. C. Cho and S. Choi, ``Nonnegative features of spectro-temporal
                                   sounds for classiﬁcation," \emph{Pattern Recognition Letters}, vol. 26 (9),
                                   2005.
                                \bibitem{zubair} S. Zubair, F. Yan, W. Wang
                               ``Dictionary learning based sparse coefficients for audio classification with max and average pooling," \emph{Elsevier Digital Signal Processing }, vol. 23, issue. 3, 2013. 
                               
                                    \bibitem{nikunen} J. Nikunen and T. Virtanen, ``Object-based audio coding using non-negative matrix factorization for the spectrogram representation," \emph{Proceedings of the 128th Audio Engineering Society Convention}, London, UK, 2010.
                               \bibitem{plumbley} M. D. Plumbley, T. Blumensath, L. Daudet, R. Gribonval, and M. E. Davies, ``Sparse representations in audio and music: from coding to source separation," \emph{Proceedings of the IEEE,} vol. 98 (6), pp. 995-1005, 2009.
                               
                          
                     \bibitem{coates} A. Coates, and Andrew Y. Ng, ``Learning Feature Representations with K-Means,"  \emph{Lecture Notes in Computer Science, Neural Networks: Tricks of the Trade}, vol. 7700, pp. 561-580, 2012.
                               
                                     \bibitem{delgado}K. Kreutz-Delgado, J. Murray, D. Rao, K. Engan, T. Lee, and T. Sejnowski, ``Dictionary learning algorithms for sparse representations," \emph{Neural Computation}, vol. 15, pp. 349-396, 2003.
                                
                                \bibitem{olshausen} B. Olshausen and D. Field, ``Emergence of simple-cell receptive field                                 properties by learning a sparse code for natural images," \emph{Nature}, vol.                                381, pp. 607-609, 1996.
                               
                               \bibitem{lewicki}M. S. Lewicki and T. J. Sejnowski, ``Learning overcomplete representations," \emph{Neural Computation}, vol. 12, pp. 337-365, 2000.
                                          
                         \bibitem{engan}K. Engan, S. O. Aase, and J. H. Husoy, ``Multi-frame compression: Theory and design," \emph{EURASIP Sig. Process.}, vol. 80, no. 10, pp. 2121-2140, 2000.            
         
%

                           \bibitem{pati} Y. Pati, R. Rezaiifar, and P. Krishnaprasad, ``Orthogonal matching pursuit: recursive function approximation with applications to wavelet decomposition," \emph{Proceedings of Asilomar Conference on Signals,    Systems and Computers}, 1993.
                                                          
                                                             \bibitem{gorodnitsky} I. F. Gorodnitsky, and B. D. Rao, “Sparse signal reconstruction from  limited data using FOCUSS: A re-weighted norm minimization algorithm,” \emph{IEEE Trans. Sig. Process.}, vol. 45, pp. 600-616, 1997.   
                                               
                     
                 
                      
                     \bibitem{skretting}K. Skretting, and K. Engan, ``Recursive least squares dictionary learning algorithm," \emph{IEEE Trans. Sig. Process.}, vol. 58, pp. 2121-2130, 2010.
                      
                      \bibitem{aharon} M. Aharon, M. Elad, and A. Bruckstein, ``K-SVD: An algorithm for designing overcomplete dictionaries for sparse representations," \emph{IEEE Trans. Sig. Process.}, vol. 54, pp. 4311-4322, 2006.
                      
                      \bibitem{dai} W. Dai, T. Xu, and W. Wang, "Simultaneous Codeword Optimization (SimCO) for dictionary update and learning," \emph{IEEE Trans. Sig. Process.}, vol. 60, no. 12, pp. 6340-6353, 2012.
                      
                      \bibitem{jafari} M. G. Jafari, M. D. Plumbley, ``Fast dictionary learning for sparse representations of speech signals," \emph{IEEE Journal. Selected Topics Sig. Process.}, vol. 5, pp.1025-1031, 2011.
                       \bibitem{kong} S. Kong, and D. Wang, `` A dictionary learning approach for classification: separating the particularity and the commonality," \emph{Lecture Notes in Computer Science, Computer Vision}, vol. 7572, pp. 186-199, 2012.
                                  
                                  \bibitem{shafiee}            S. Shafiee, F. Kamangar andV. Athitsos, and J. Huang,           ``The role of dictionary learning on sparse representation-based classification," \emph{Proc. Int. Conf. Pervasive Technologies Related to Assistive Environments}, no. 47, 2013.  
                      
                    \bibitem{mallat}S. G. Mallat, and Z. Zhang, ``Matching pursuits with time-frequency dictionaries," \emph{IEEE Trans. Sig. Process.}, vol. 41, pp. 3397-3415, 1993.
                         
                       
                            
                            \bibitem{chen} S. S. Chen, D. L. Donoho, and M. A. Saunders, ``Atomic decomposition by basis pursuit," \emph{SIAM Rev.}, vol. 43 (1), pp. 129-159, 2001.
                                 
           
 
           
                    \bibitem{kv} K V Vijay Girish, A G Ramakrishnan and T V Ananthapadmanabha, ``Hierarchical classification of speaker and background noise and estimation of SNR using sparse representation," \textit{To be presented in INTERSPEECH, 2016 }
                    
           
                  \bibitem{kv1} K V Vijay Girish, T V Ananthapadmanabha and A G Ramakrishnan, ``A dictionary learning and source recovery based approach to classify diverse audio sources," \textit{arXiv:1510.07774 [cs.SD]
                  }, 2015
                  
                           \bibitem{spath} H. Spath, ``Cluster Dissection
                           and Analysis: Theory," \textit{FORTRAN Programs, Examples. Translated
                           by J. Goldschmidt.} New York: Halsted Press, 1985.
                           
                           \bibitem{park} H.S. Park and C.H. Jun, ``A simple and fast algorithm for K-medoids clustering," \textit{ Expert Systems with Applications}, vol. 36 (2), pp. 3336–3341, 2009.
                           
                           
                         
       \bibitem{fu} Y. Fu, H. Li, Q. Zhang and J. Zou, ``Block-sparse recovery via redundant block OMP," \textit{Signal Processing, Elsevier}, vol. 97, pp. 162--171,  2014.
       
       
       \bibitem{noisex} Noisex-92. [Online], Available:
           \textit{http://www.speech.cs.cmu.edu/}\\\textit{comp.speech/Section1/Data/noisex.html}
       
                \bibitem{loizou} P. C. Loizou, ``Speech Enhancement Based on Perceptually Motivated Bayesian Estimators of the Magnitude Spectrum," \textit{ IEEE Transactions on Speech and Audio Processing}, vol. 13, pp. 857--869, 2005.
%
%

         
         
           
           \bibitem{tropp} J. A. Tropp,   ``Greed is good: algorithmic results for sparse approximation," \emph{ IEEE Trans. Info. Theory}, vol. 50, pp. 2231-2242, 2004.
           
           \bibitem{vincent} E. Vincent, R. Gribonval and C. Fevotte, ``Performance Measurement in           Blind Audio Source Separation," \emph{IEEE Trans. Audio, Speech, and Lang. Process.}, vol. 14, pp. 1462 - 1469, 2006.
           
         
       \bibitem{rose} R. C. Rose,  E. M. Hofstetter and D. A. Reynolds Integrated models of signal and background with application to speaker identification in noise," \textit{IEEE Transactions on Speech and Audio Processing }, vol. 2, pp. 245--257, 1994.
         
      \bibitem{mohammadiha} N Mohammadiha, P. Smaragdis and A. Leijon,   ``Supervised and Unsupervised Speech Enhancement Using Nonnegative Matrix Factorization,"  \textit{ IEEE Transactions on Audio, Speech and Language Processing}, vol. 21, pp. 2140--2151, 2013.
           
    \end{thebibliography}
\end{document}